\pdfoutput=1
\documentclass[a4paper,american,floatfix,pdftex,superscriptaddress,twoside,%
aps,prl,%
reprint,
]{revtex4-2}%
\usepackage[T1]{fontenc}
\usepackage{multirow}
\usepackage{graphicx}%
\usepackage[utf8]{inputenc}
\usepackage{hyperref, hypernat}
\usepackage[ams,todos]{CMImacros}
\usepackage[boxed]{algorithm2e} % this should go before hyperref (or in alphabetical %order) but
\graphicspath{{./figures/}} % define figure directory
\usepackage{tabularx}
\usepackage{longtable}
\usepackage{booktabs}
\newcommand{\cfeldesy}{\affiliation{Center for Free-Electron Laser Science CFEL, Deutsches
      Elektronen-Synchrotron DESY, Notkestr. 85, 22607 Hamburg, Germany}}%
\newcommand{\uhhcui}{\affiliation{Center for Ultrafast Imaging, Universität Hamburg, Luruper
      Chaussee 149, 22761 Hamburg, Germany}}%
\newcommand{\uhhphys}{\affiliation{Department of Physics, Universität Hamburg, Luruper Chaussee 149,
      22761 Hamburg, Germany}}%
\newcommand{\uhhmaths}{\affiliation{Department of Mathematics, Universität Hamburg, Bundesstr. 55,
      20146, Hamburg, Germany}}%
\newcommand{\ysemail}{\email[Email:~]{yahya.saleh@uni-hamburg.de}}%
\newcommand{\ayemail}{\email[Email:~]{andrey.yachmenev@robochimps.com}}%
\newcommand{\cmiweb}{\homepage[\\ URL:~]{https://www.controlled-molecule-imaging.org}}%

\SetAlCapSkip{\abovecaptionskip}

\begin{document}
\title{Computing excited states of molecules using normalizing flows}%
\author{Yahya Saleh}\ysemail\cmiweb\uhhmaths\cfeldesy %
\author{Álvaro Fernández Corral}\cfeldesy\uhhphys %
\author{Emil Vogt}\cfeldesy %
\author{Armin Iske}\uhhmaths%
\author{Jochen Küpper}\cfeldesy\uhhphys\uhhcui%
\author{Andrey Yachmenev}\ayemail\cfeldesy\uhhcui%
\date{\today}

\begin{abstract}
	Calculations of highly excited and delocalized molecular vibrational states
	are computationally challenging tasks, which strongly depends on the choice
	of coordinates for describing vibrational motions. We introduce a new method
	that leverages normalizing flows -- parametrized invertible functions -- to
	learn optimal vibrational coordinates that satisfy the variational principle. This approach
	produces coordinates tailored to the vibrational problem at hand,
	significantly increasing the accuracy and enhancing basis-set convergence of the
	calculated energy spectrum. The efficiency of the method is demonstrated in
	calculations of the 100 lowest excited vibrational states of H$_2$S, H$_2$CO,
	and HCN/HNC. The method effectively captures the essential vibrational
	behavior of molecules by enhancing the separability of the Hamiltonian and
	hence allows for an effective assignment of approximate quantum numbers. We
	demonstrate that the optimized coordinates are transferable across different
	levels of basis-set truncation, enabling a cost-efficient protocol for
	computing vibrational spectra of high-dimensional systems.
 
\end{abstract}

\maketitle
\section{Introduction}
Accurate calculations of highly excited vibrational states of polyatomic
molecules are essential for unravelling increasingly rich experimental
spectroscopic information and understanding the dynamics of intermolecular
motions. The highly excited molecular vibrations are especially important in
fields such as chemical reactivity~\cite{Li:Science383:746,
   Auerbach:NatSci1:e10005, Foley:Science374:1122} and
collisions~\cite{Margulis:Science380:77, Rahinov:PCCP26:15090}, relaxation
processes~\cite{Meng:JPCC126:12003} and stimulated
emission~\cite{Yoneda:JPCA127:5276}, as well as spectroscopic probing of
high-temperature environments found on exoplanets~\cite{Wright:MNRAS512:2911,
   Wright:AJ166:41} and in industrial
applications~\cite{Pinkowski:MeasSciTechnol31:055501, Ehn:ApplSpec71:341}.

A range of variational and perturbative methods were developed for predicting
vibrational spectra of molecules~\cite{Yurchenko:JMS245:126,
   Matyus:JCP130:134112, Meyer:MCTDH:2009, Bowman:IRPC22:533,
   Christiansen:JCP120:2140, Hansen:JCTC6:235, Christiansen:JCP120:2149}. These
methods solve the eigenvalue problem for a vibrational Hamiltonian, which is
constructed using appropriately chosen vibrational coordinates. The choice of
coordinates is a crucial task that directly affects the accuracy of energy
calculations. When using a direct product basis of univariate functions, a key
challenge is selecting coordinates that provide a large degree of separability
of vibrational motions, thereby reducing the computational effort required to
solve the vibrational eigenvalue problem~\cite{Bramley:MP73:1183}. This is
particularly important for calculations of delocalized vibrational states of
floppy molecules~\cite{Matyus:ChemComm59:366, Bacic:ARPC40:469}, such as van der
Waals complexes~\cite{Hutson:ARPC41:123}, molecules near
dissociation~\cite{Simko:ACIE62:e202306744}, or high-energy excitations in
general~\cite{Yachmenev:JCP143:014105}, where couplings between different
vibrational modes are prominent.

Rectilinear normal coordinates provide a natural starting point for seeking
separability in vibrational problems. However, they become less effective for
highly excited states and are generally not suited for floppy molecules, \eg,
weakly bound complexes, which naturally sample configurations far from their
reference equilibrium geometry. Alternative curvilinear coordinate systems, such
as Radau~\cite{Wang:JCP129:234102}, Jacobi~\cite{Gatti:JCP108:8804,
   Leforestier:JCP114:2099}, valence~\cite{Bramley:JCP98:1378}, and
polyspherical~\cite{Iung:IJQC106:130, Gatti:PhysRep484:1, Klinting:JCTC16:4505}
coordinates, were successfully applied in the vibrational calculations of
various floppy polyatomic molecules~\cite{Bowman:MP106:2145,
   Matyus:JCP130:134112}. Choosing the optimal coordinates requires a
combination of intuition, consideration of the symmetries of the system, and
prior knowledge of the potential energy landscape. This task is particularly
challenging for floppy molecules and generally large systems. Several general
strategies were recently developed to guide the selection and design of
vibrational coordinates, drawing from the available pool of known curvilinear
and rectilinear coordinates~\cite{Oenen:JCP160:014104, MendiveTapia:JCTC19:1144,
   Schneider:JCP161:094102}.

Due to the diversity of nuclear motions and their dependence on molecular size
and bonding topology, no single coordinate system is universally optimal for
describing the vibrations of different molecules. One promising approach to
improve the effectiveness of a coordinate system involves developing general
coordinates, parametrized by variables that can be optimized to minimize
vibrational couplings or energy levels. Such general coordinates, expressed as
linear combinations of normal coordinates~\cite{Thompson:JCP77:3031, Mayrhofer:TCA92:107},
curvilinear coordinates~\cite{Zuniga:JCP122:224319, Zuniga:JCP115:139, Bulik:JCP147:044110,
Bowman:JCP90:2708, Bacic:JCP90:3606, Yagi:JCP137:204118,
Thomsen:JCP140:154102}, or as a quadratic function of normal
coordinates~\cite{Chan:JCTC8:6}, were shown to significantly enhance the
accuracy of variational calculations. Despite these developments, the broader
application of coordinate optimization in variational calculations remains
largely unexplored, with previous efforts generally limited to linear parametrizations
specific to particular systems.

In this work, we introduce a new general nonlinear parametrization for
vibrational coordinates that is based on normalizing
flows~\cite{Papamakarios:JMLR22:1}, implemented using a neural network. The
parameters of the neural network are optimized using the variational principle.
Applied to the calculation of the 100 lowest vibrational states in H$_2$S,
H$_2$CO, and HCN/HNC molecules, the present approach achieves several orders of
magnitude greater accuracy in energy predictions compared with commonly used
curvilinear coordinates for the same number of basis functions. The optimized vibrational coordinates
effectively capture the underlying physics of the problem, reduce couplings
between different vibrational modes, and remain consistent across various levels
of basis-set truncation. Building on this property, we propose a cost-efficient
approach in which the coordinates are first optimized using a small number of
basis functions
and then applied to calculations with a larger number of basis functions, keeping the parameters
fixed.

\section{Methods}

\subsection{Enhancing a basis by change of coordinates}

We begin by choosing a truncated set of orthonormal basis functions
$\{\phi_n(\mathbf r)\}_{n=0}^N$ of $L^2$ along with an invertible map $g_\theta$
parametrized by a set of parameters $\theta$. $g_\theta$ maps an initial set of
vibrational coordinates $\mathbf r$ to a new set of coordinates $\mathbf q$ of
the same dimension, \ie, ${\mathbf q}=g_\theta(\mathbf r)$. To improve the
approximation properties of the basis functions $\phi_n$, we evaluate them in
$\mathbf q$
%we compose them with the mapping $g_\theta$
to obtain a new set of augmented basis functions $\{\gamma_n({\mathbf q};\theta)\}_{n=0}^N$ defined as
\begin{align}\label{eq:gamma_n}
  \gamma_n({\mathbf q};\theta) := \phi_n\bigl(g_\theta(\mathbf{r})\bigr)\sqrt{|\det \nabla_{\mathbf r} g_\theta(\mathbf r)|},
\end{align}
where multiplying by the inverse of the square root of the determinant of the
Jacobian ensures that the basis functions remain orthonormal with respect to the
$L^2$-inner product, independent of the values of $\theta$. Inducing augmented
basis functions by a nonlinear change of variables is analogous to inducing an
augmented probability distribution $p$ from a base distribution $p_0$ using a
change of variables $g_\theta$, commonly referred to as a normalizing flow in
the machine learning literature~\cite{Rezende:ICML37:1530,
   Papamakarios:JMLR22:1}. Therefore, we refer to $g_\theta$ as a normalizing
flow and to $\mathbf q$ as normalizing-flow coordinates.

The map $g_\theta$ can, in principle, be any differentiable invertible function.
However, to maintain the completeness of the augmented basis set
$\{\gamma_n({\mathbf q};\theta)\}_{n=0}^{\infty}$, the normalizing flow must be
bijective~\cite{Saleh:arXiv2406:18613}. We construct $g_\theta$ using an invertible residual neural network
(iResNet)~\cite{Behrmann:ICML2019:573}. We refer to the supplementary information for a more
detailed explanation of the equivalence between optimizing basis sets and
vibrational coordinates.

\subsection{Architecture}
By construction, the iResNet, which is commonly used for image processing,
places no restrictions on the image domain. However, in many computational
physics and chemistry applications, the domain of internal coordinates is
inherently bounded. For example, in vibrational calculations, internal
coordinates often represent distances, which are strictly positive, or angles,
which are typically periodic or confined to a finite interval such as $[0,\pi]$ or $[0,2 \pi]$.
Mapping into non-physical or redundant ranges of internal coordinates can lead
to inaccurate numerical outcomes and violations of the variational limit. For
example, the iResNet output may extend into coordinate regions where the
potential is undefined, leading to unreliable results.

\begin{figure}[b]
   \includegraphics[width=\linewidth]{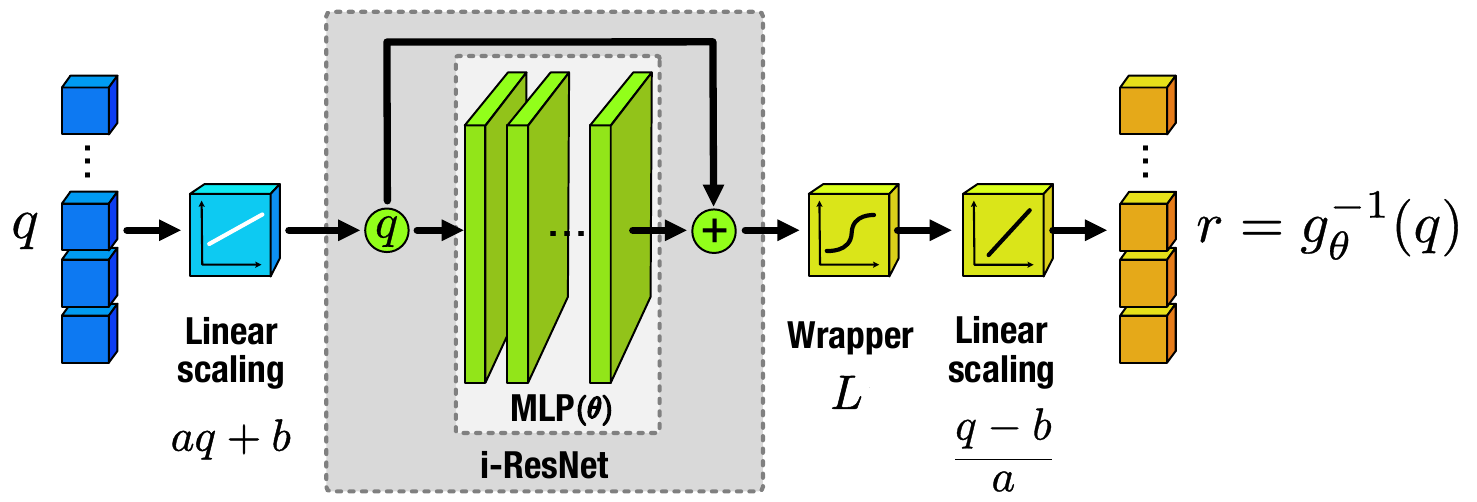}
   \caption{\textbf{Schematic diagram of the computational workflow of the
         normalizing-flow function.} As wrapping functions we used $L=\tanh(x)$.
      A fixed scaling procedure was applied to map the initial coordinate ranges
      to an identical domain. The linear scaling parameters $a$ and $b$ were
      optimized together with the MLP $\theta$-parameters.}
   \label{fig:iresnet}
\end{figure}

To address these issues, we developed an invertible flow that enables control over the
output ranges. First, we mapped the ranges of the quadratures of the basis sets
in each dimension to $[-1, 1]$ using a fixed linear scaling. This way, we allow
any domain of the primitive basis set to be handled identically. Next, the
iResNet was applied, producing an unbounded output. To map the output back to a
finite interval, we applied a wrapper function, $L$. This wrapper function can
be any mapping that transforms an infinite domain into a finite domain. We
selected the $\tanh$ function to map values to the interval $[-1, 1]$. Finally,
we applied a linear scaling $\mathbf{a} \cdot \mathbf{x} + \mathbf{b}$ to adjust
the output to the desired target interval. The scaling parameters $\mathbf{a}$
and $\mathbf{b}$ were optimized as part of the workflow. To ensure the output
remains within the correct interval, we used a subparametrization of the linear
parameters $a_i(\alpha), b_i(\beta)$, where $\alpha$ and $\beta$ are optimizable
variables. We applied different functions for different coordinates, depending
on their specific range requirements. Three different subparametrizations were
used to accommodate infinite, semi-infinite, and finite intervals. A schematic
of this workflow is provided in \autoref{fig:iresnet}.

\subsection{Construction of the Hamiltonian}
The matrix elements of the vibrational kinetic and potential energy operators in
the augmented basis \eqref{eq:gamma_n} can be expressed by introducing a change
of variables ${\mathbf q}=g_\theta(\mathbf r)$ in the integrals. For the
potential, this results in the following expression
\begin{eqnarray}\label{eq:V_n'n}
   \mathbf{V}_{n'n}=\langle \gamma_{n'}|V|\gamma_{n} \rangle = \int \phi_{n'}^*(\mathbf q) V(g_\theta^{-1}(\mathbf q)) \phi_n(\mathbf q) \mathrm{d}\mathbf{q}.
\end{eqnarray}
This illustrates that the normalizing flow $g_\theta$ effectively modifies the
coordinates in which the operators are expressed for a fixed set of basis
functions $\{\phi_n\}_n$.

The matrix elements of the kinetic energy operator in the augmented basis set
are given by
\begin{align}\label{eq:keo}
	&T_{nn'} = \int \phi_n^{*}(\mathbf{q}) \hat{T}(g_\theta^{-1}(\mathbf{q})) \phi_{n'}(\boldsymbol q) \mathrm{d}\mathbf{q} & \\
	&= \frac{\hbar^2}{2} \sum_{kl} \int \left[ \left(\frac{1}{2\sqrt{D}}\frac{\partial D}{\partial q_k} + \sqrt{D}\frac{\partial}{\partial q_k}\right) \phi_n^*(\mathbf{q})\right]& \nonumber \\
	&\times\sum_{\lambda\mu}\frac{\partial q_k}{\partial r_\lambda}G_{\lambda\mu}(g_\theta^{-1}(\mathbf{q}))\frac{\partial q_l}{\partial r_\mu}& \nonumber \\
	&\times\left[ \left(\frac{1}{2\sqrt{D}}\frac{\partial D}{\partial q_l} + \sqrt{D}\frac{\partial}{\partial q_l}\right) \phi_{n'}(\mathbf{q})\right] \mathrm{d}\mathbf{q}.& \nonumber
\end{align}
$D=\abs{1/\det \nabla_{\mathbf q} g_\theta^{-1}(\mathbf q)}$, $G_{\lambda\mu}$
is the kinetic-energy matrix, $\lambda$, $\mu$ indices are used to denote the
elements of the coordinate vector $\mathbf{r}$, and $k$, $l$ indices denote the
elements of the coordinate vector $\mathbf{q}$, \ie,
$q_k=g_{\theta,k}(\mathbf{r})$ and
$r_\lambda=g_{\theta,\lambda}^{-1}(\mathbf{q})$. The differential operators,
$\frac{\partial}{\partial q_k}$, only operate inside the square brackets. To
obtain this formula, we employed integration by parts, enabling the second-order
derivative operator to act symmetrically on both the bra and ket functions as
first-order derivatives. The boundary term is omitted as its contribution is
zero for most integration domains. Additionally, the kinetic energy operator
includes the so-called pseudopotential term, which originates from the
transformation from Cartesian to the initial internal coordinates. It is calculated
as
\begin{align}\label{eq:pseudo}
U=
\frac{\hbar^2}{32} \sum_{\lambda}\sum_{\mu}
\frac{G_{\lambda\mu}}{\tilde{g}^2} \frac{\partial\tilde{g}}{\partial r_\lambda}
\frac{\partial\tilde{g}}{\partial r_\mu}
+4\frac{\partial}{\partial r_\lambda}
\Bigg(
\frac{G_{\lambda\mu}}{\tilde{g}}
\frac{\partial\tilde{g}}{\partial r_\mu}
\Bigg),
\end{align}
where $\tilde{g} = \det \left( G^{-1} \right)$. The pseudopotential is a scalar
operator and its matrix elements in the transformed basis can be expressed
analogously to the potential energy matrix elements in \eqref{eq:V_n'n}.

\subsection{Optimization}
We approximate the vibrational wavefunctions $\Psi_m$ $(m=1..M)$ as linear
combination of augmented basis functions \eqref{eq:gamma_n}, \ie,
\begin{align}\label{eq:Psi_m}
	\Psi_m({\mathbf q}) \approx \sum_{n\leq N} c_{nm}\gamma_n({\mathbf q}; \theta).
\end{align}
The linear-expansion coefficients $c_{nm}$ and the normalizing-flow parameters
$\theta$ are determined using the variational principle, by minimizing the
energies of the ground and excited vibrational states. For the coefficients,
this is equivalent to solving the eigenvalue problem
$\mathbf{E} = \mathbf{C}^{-1}\mathbf{H}\mathbf{C}$, where
$\mathbf{C} = \{c_{nm}\}_{n,m}^{N,M}$, $\mathbf{E}=\{E_m\}_m^M$ are the
vibrational energies, and $\mathbf{H}=\mathbf{T}+\mathbf{V}$ is the sum of
matrix representations of the kinetic and potential energy operators, given by \eqref{eq:V_n'n} and \eqref{eq:keo}.

\begin{figure*}
   \includegraphics[width=\linewidth]{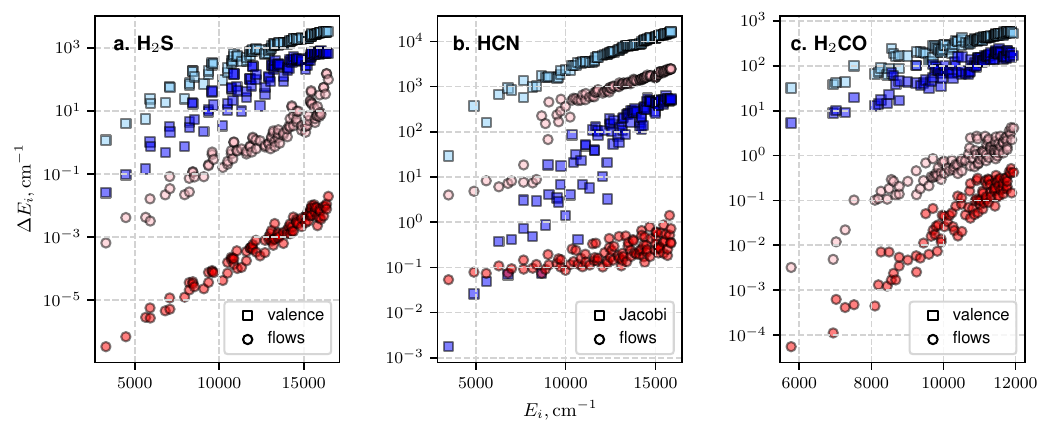}
   \caption{\textbf{Convergence of H$_2$S, HCN and H$_2$CO vibrational energy
         levels.} Plotted are the 100 lowest energy discrepancies using standard (blue
      squares) and normalizing-flows (red circles) coordinates. Light and dark
      colors represent truncations corresponding to a smaller and larger number
      of basis functions, respectively. \textbf{a} Energy discrepancies for
      H$_2$S at $P_\text{max}=12$ (140 basis functions) and $20$ (506).
      \textbf{b} Energy discrepancies for HCN at $P_\text{max}=12$ (140) and
      $32$ (1785). \textbf{c} Energy discrepancies for H$_2$CO at
      $P_\text{max}=9$ (1176) and $12$ (3906).}
   \label{fig:convergence}
\end{figure*}

Because vibrational energies are nonlinear functions of the parameters $\theta$,
these parameters are optimized using gradient descent methods. The optimization
is guided by a loss function derived from the variational principle and may
involve minimizing quantities such as the sum of vibrational energies, the trace
of the Hamiltonian matrix, or the matrix exponential. A loss function expressed
as the sum of all energies spanned by the chosen basis set is equivalent to the
trace of the Hamiltonian matrix,
\begin{align} \label{eq:L_theta}
	\mathcal{L}_\theta = \sum_{n\leq N}E_n = \Tr(\mathbf{H}) \rightarrow \min_\theta.
\end{align}
This loss function has a relatively low computational cost, as it decouples the
nonlinear parameters $\theta$ from the eigenvector coefficients $c_{nm}$,
requiring only the evaluation of diagonal elements of the Hamiltonian matrix
when the initial basis is orthonormal. In contrast, when the loss function is
based on the sum of a subset of the lowest energies, the parameters depend on
the eigenvector coefficients $c_{nm}$, and a repeated solution of the eigenvalue
problem during optimization is required. Despite the added complexity, the high
accuracy achieved even with small number of basis functions can potentially outweigh the computational
costs of the repeated matrix diagonalization. In our calculations we used
the sum of a subset of all vibrational energies as the loss function, \emph{vide infra}. A cost-efficient optimization and application strategy that mitigates the cost of repeated matrix diagonalization is discussed in \autoref{sec:transferability}.

The evaluation of the matrix elements in \eqref{eq:V_n'n} and \eqref{eq:keo} is one of the most computationally
demanding parts. In this work, we employed Gaussian quadratures to compute the
necessary integrals, altering the quadrature degree at different optimization
steps to prevent overfitting. We found that alternating between smaller
quadratures during optimization was computationally more efficient while still
converging to the same values of the parameters $\theta$ as those obtained using
a larger quadrature. After convergence, the final energies and wavefunctions
were computed by solving the eigenvalue problem with a large quadrature for
accurate integral evaluations. For higher-dimensional systems, more efficient
techniques such as sparse-grid methods~\cite{Avila:JCP139:134114} or
collocation~\cite{Yang:CPL153:98} can be used. Alternatively, Monte-Carlo
methods~\cite{Toulouse:JCP128:174101, Cuzzocrea:JCTC16:4203} may be employed
when high accuracy is not required.

\subsection{Computational details}
The accuracy and performance of our approach were validated in calculations of
vibrational states for hydrogen sulfide H$_2$S, formaldehyde H$_2$CO, and
hydrogen cyanide/hydrogen isocyanide HCN/HNC isomers.
For H$_2$S and H$_2$CO, we used valence coordinates as the reference
coordinates and employed a direct product of Hermite functions as the basis set.
For H$_2$S, the direct product basis was constructed by considering only combinations of
one-dimensional vibrational quantum numbers $(n_1,n_2, n_3)$ that satisfy the
polyad condition $2n_{1} + 2n_{2} + n_{3} \leq P_\text{max}$, where $n_1$,
$n_2$, and $n_3$ correspond to the vibrational quanta for $r_{\text{SH}_1}$,
$r_{\text{SH}_2}$, and $\alpha_{\angle\text{H}_1\text{SH}_2}$ valence
coordinates, respectively. For H$_2$CO, we applied the basis truncation
condition $2n_{1}+ 2n_{2} + 2n_{3}+ n_{4}+ n_{5} + n_{6}\leq P_\text{max}$,
where $n_1,\dots,n_6$ correspond to the vibrational quanta for valence
coordinates $r_{\text{CO}}$, $r_{\text{CH}_1}$, $r_{\text{CH}_2}$,
$\alpha_{\angle\text{O}\text{CH}_1}$, $\alpha_{\angle\text{O}\text{CH}_2}$, and
$\tau$, the dihedral angle between the OCH$_1$ and OCH$_2$ planes. For HCN/HNC, we used the basis truncation condition $2n_{1} + 2n_{2} + n_{3} \leq P_\text{max}$ and Jacobi reference coordinates
$r_\text{CN},R,\alpha_{\angle \text{R-CN}}$, where $R$ is the distance between
the hydrogen atom and the center of mass of the C--N bond, and
$\alpha_{\angle\text{R-CN}}$ is the angle between these coordinate vectors. Hermite functions were used for the two radial coordinates and Legendre functions for
the angular coordinate. The Legendre functions were multiplied by
$\sin^{1/2}(\alpha_{\angle \text{R-CN}})$ to ensure the correct behaviour of the
wavefunction at the linear geometry of the molecule, where the Hamiltonian
becomes singular~\cite{Chubb:JCP149:014101,Yurchenko:JCP153:154106}. For all
molecules, we employed spectroscopically refined potential energy surfaces
(PES)~\cite{Azzam:MNRAS460:4063,Al-Refaie:MNRAS448:1704,Mourik:JCP115:3706} and
numerically constructed exact kinetic energy operator using the method described
in~\cite{Yachmenev:JCP143:014105, Matyus:JCP130:134112}.

To model the normalizing flow $g_\theta$, we used an iResNet consisting of 10
blocks. Each block was represented by a dense neural network comprising two
hidden layers with unit sizes $[8,8]$ and an output layer of $n$ units, where
$n$ corresponds to the number of coordinates. A more detailed description of the
normalizing-flow architecture is available in the supplementary information. The
normalizing-flow parameters were optimized variationally by minimizing the sum
of the 100 or 200 lowest vibrational energies. Generally, 1000 iterations were
enough to achieve good convergence. Benchmark energies were computed
with basis sets truncated at $P_\text{max}=60$ (optimized for $P_\text{max}=12$)
for H$_2$S, $P_\text{max}=16$ (optimized for $P_\text{max}=9$) for H$_2$CO, and
$P_\text{max}=44$ for HCN/HNC.

\section{Results}
\subsection{Computed Vibrational Energies}
On average, over the 100 lowest energies, the calculations converged with an
accuracy of $0.04$~\invcm for H$_2$S, $0.53$~\invcm for H$_2$CO, and $0.03$
\invcm for HCN/HNC compared to the reference values reported in the
literature~\cite{Azzam:MNRAS460:4063, Mourik:JCP115:3706,
   Al-Refaie:MNRAS448:1704}. We thus considered our results to be converged and
used them as benchmark data throughout the rest of the manuscript. A table
summarizing the deviations of vibrational energies from the reference values is
provided in the supplementary information.

The absolute error for the 100 lowest vibrational states of H$_2$S, H$_2$CO, and
HCN/HNC, as a function of the basis-set truncation parameter $P_\text{max}$, is
shown in \autoref{fig:convergence}. For each molecule, the results of two variational
calculations are presented, one using reference valence or Jacobi coordinates
and another using the optimized normalizing-flow coordinates. With the same
number of basis functions, coordinate
optimization resulted in up to five orders of magnitude improvement in the
accuracy of vibrational energy calculations compared to using the standard
reference coordinates. Extrapolating to a larger number of basis functions, we estimate that
matching the same accuracy using the reference coordinates would require
approximately an order of magnitude increase in the number of basis functions.

We note that the convergence of results can also be improved by increasing the
complexity of the normalizing-flow function. A detailed analysis of this effect, along with an investigation of how varying the number of target states affect the normalizing-flow coordinates, is
provided in the supplementary information.

Direct product basis sets can be improved using basis-set contraction, which
involves partitioning the total Hamiltonian into subsystems, solving
reduced-dimensional variational problems for each, and then using these
solutions for the full-dimensional problem~\cite{Wang:JCP117:6923,
   Felker:JCP151:024305}. For molecules like H$_2$S and H$_2$CO, basis set
contraction works well due to near-separability of valence coordinates in the PES. However, this approach becomes more challenging for
floppy molecules like HCN/HNC, where two of the Jacobi coordinates
are strongly coupled.

\begin{figure}
   \includegraphics[width=1.\linewidth]{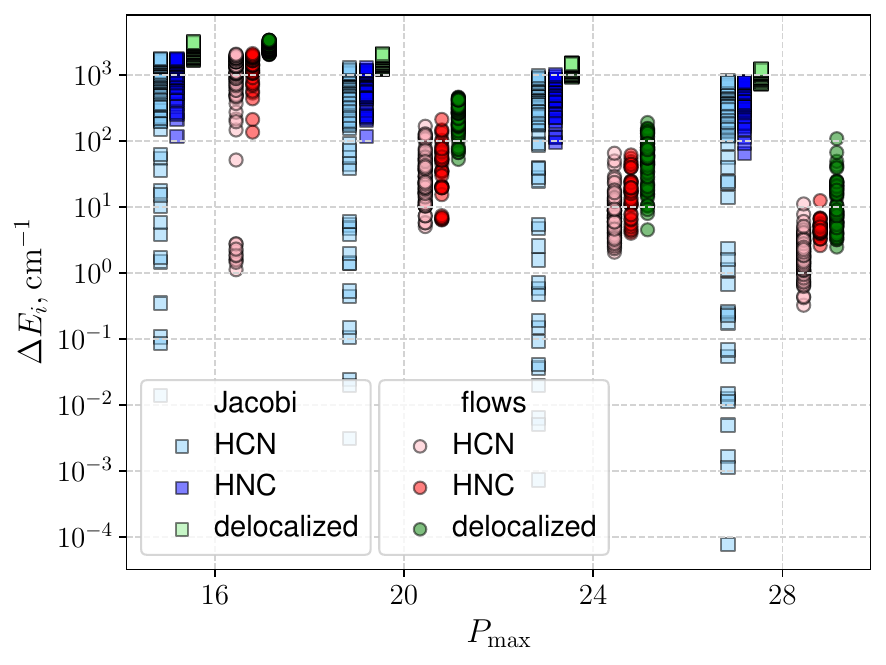}
   \caption{\textbf{Convergence of HCN/HNC vibrational energy levels.} Shown are
      the lowest 200 energies for using Jacobi coordinates (squares) and
      normalizing-flow coordinates (circles). The energy discrepancies
      ($\Delta E_i$) relative to our converged benchmark reference are shown for
      several basis sets, truncated at $P_\text{max}=12$ (140 basis functions),
      16 (285), 20 (506), 24 (811), and 28 (1200). Vibrational states assigned
      to the HCN isomer, the HNC isomer, and states with an energy above the
      isomerization barrier (delocalized) are differentiated by color. All
      states are slightly offset along the $P_\text{max}$ axis for visual
      clarity. }
   \label{fig:isomers}
\end{figure}
This challenge is illustrated in \autoref{fig:isomers} for the first 200
vibrational energies of HCN/HNC. The figure presents results of basis set
contraction using Jacobi coordinates alongside those obtained with a direct
product basis set of primitive functions, \ie, Hermite and Legendre, using
optimized normalizing-flow coordinates. The contracted basis was constructed by
partitioning the Hamiltonian into one-dimensional subsystems for each of the
Jacobi coordinates, with the HCN isomer equilibrium geometry as the reference
configuration. As shown in \autoref{fig:isomers}, the contracted basis
significantly improves convergence compared to the product basis set results in
\autoref[b]{fig:convergence}, but not for all vibrational states.
High accuracy is only achieved for states localized around the HCN minimum of
the PES, while delocalized states and states localized around the HNC minimum
show little improvement. In contrast, the optimized normalizing-flow coordinates
provide a balanced description of all localized and delocalized states, with a
much smaller spread in errors across different states.

\subsection{Intepretability}
\begin{figure}
   \includegraphics[width=\columnwidth]{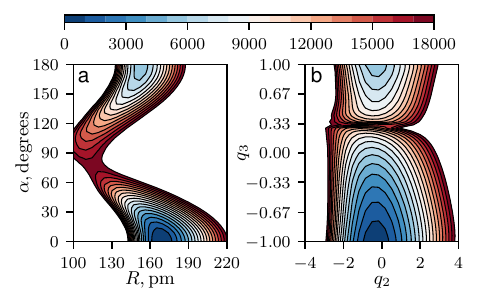}
   \caption{\textbf{Two-dimensional cuts of the HCN/HNC potential energy
         surface.} \textbf{a} Cut along the Jacobi coordinates $R$ and
      $\alpha_{\angle \text{R-CN}}$. \textbf{b} Cut along the optimized
      normalizing-flow coordinates. The optimization was performed for a basis
      set truncated at $P_\text{max}=16$ (285 basis functions) with the loss
      function defined as the sum of the 100 lowest vibrational state energies.
      An effective decoupling of the surface in the normalizing-flows
      coordinates is patent. }
   \label{fig:PES}
\end{figure}
To gain insight into the interpretation of the optimized normalizing-flow
coordinates, we plotted in \autoref{fig:PES} the two-dimensional cut of the PES
of HCN/HNC along the strongly coupled Jacobi coordinates, $R$ and
$\alpha_{\angle \text{R-CN}}$, \ie, $V(r_0, R, \alpha_{\angle\text{R-CN}})$,
alongside the corresponding cut in the optimized normalizing-flow coordinates,
$V(g_\theta^{-1}(q_1, q_2, q_3))$. In these plots, $r_\text{CN}$ coordinate is
fixed at its equilibrium value $r_0$ and $q_1=0$. The potential is clearly
highly anisotropic when expressed in Jacobi coordinates (panel a), which leads
to a strong coupling between the two vibrations. In contrast, when expressed in
the optimized coordinates (panel b), the HCN$\leftrightarrow$HNC minimum energy
isomerization pathway is practically a straight line along the coordinate $q_3$
at $q_2\approx0$. This reduction in anisotropy explains why coordinate
optimization improves the convergence in the product basis. The optimization
achieves an effective coordinate decoupling of the PES, which allows for a
better approximation of the eigenfunctions of the Hamiltonian by the chosen
direct product basis. In addition, it is evident from the spacing between the
contour lines along the flows coordinate $q_2$, that the potential becomes more
harmonic in this dimension in comparison to $R$ in the Jacobi coordinates. The
same behavior was observed when comparing $q_1$ and $r_\text{CN}$. This is
expected, as Hermite functions - the solutions of the quantum harmonic
oscillator - were used as the basis for stretching coordinates.

\subsection{Assignment of approximate quantum numbers}
Assigning approximate quantum numbers to
computed eigenstates connects numerical results to their spectroscopic
interpretation. Typically, as the complexity of the method for solving the
Schrödinger equation increases, so does the difficulty of assignment. In many
cases, less accurate but more interpretable effective models are more practical
than highly accurate methods, as they facilitate approximate quantum number assignment and enhance the interpretability of experimental spectra.

\begin{table}[b]
   \centering
   \begin{tabular*}{\linewidth}{@{\extracolsep{\fill}}lcccc}
     \toprule
     Measure &   Median & Mean & Min  & N$_\text{assign}$\\
     \midrule
     Jacobi (HCN) &  0.27 &  0.35  & 0.05 &  24 \\
     Jacobi (HNC) &  0.17 &  0.25  & 0.04&  15 \\
     Jacobi (HCN/HNC) &  0.39 &  0.46  & 0.12&  39 \\
     Flows (HCN) & 0.80 & 0.79 &  0.40 & 97 \\
     \bottomrule
   \end{tabular*}
   \caption{\textbf{Projection-based assignment metrics for vibrational states of
      HCN/HNC}. The tabulated values are the median, mean and minimum values of
      the largest norm-square projection coefficients obtained by projecting the
      vibrational wavefunctions for 100 states onto products of one-dimensional
      eigenfunctions. In the last column, N$_\text{assign}$, the number of
      vibrational states (out of 100) that could be unambiguously assigned a
      unique set of approximate quantum numbers are shown. }
   \label{tab:assignment}
\end{table}
The enhanced coordinate decoupling in the HCN/HNC Hamiltonian suggests that
assigning approximate quantum numbers to computed eigenstates is more straightforward in
normalizing-flow coordinates compared to reference Jacobi coordinates. In
\autoref{tab:assignment}, we compare the accuracy of projection-based assignment
of approximate quantum numbers for the first 100 eigenstates of HCN/HNC. Projections were
performed onto one-dimensional eigenfunctions (a contracted basis) expressed in
either Jacobi or normalizing-flow coordinates. In the limit of convergence of
both the three- and one-dimensional eigenfunctions, the projection-based
assignment depends only on the choice of coordinates and not on the basis used
to compute the eigenfunctions. A unique assignment is ensured when the
norm-square of the largest absolute projection coefficient is larger than 0.5.
The normalizing-flow coordinates were optimized for 100 eigenstates with
$P_\text{max} = 28$.

The contracted basis was constructed by partitioning the Hamiltonian into
one-dimensional subsystems corresponding to each coordinate. For Jacobi
coordinates, either the HCN or HNC equilibrium geometry was used as reference
configurations. In contrast, for normalizing-flow coordinates, only the HCN
equilibrium geometry was needed due to reduced vibrational coupling. The
assignment of approximate quantum numbers is significantly more accurate using the
normalizing-flow coordinates with 97 uniquely assigned states compared to only
39 using Jacobi coordinates. This highlights the benefits of optimized
coordinate transformations for more reliable spectroscopic interpretations of
the results.

\subsection{Transferability}\label{sec:transferability}
We found that the converged iResNet parameters remained nearly identical when
optimized using different basis-set truncations, suggesting that there exist
unique optimal vibrational coordinates for a given type of basis. Leveraging
this finding, we developed a cost-efficient approach where the flow coordinates
are first optimized using a small number of basis functions and then applied with fixed
parameters in calculations with a larger number of basis functions. The size and computational cost of the quantities that depend on the normalizing flow, such as $\frac{\partial q_{\alpha}}{\partial r_l}$,
$\frac{\partial D}{\partial r_l}$, $\frac{1}{D}$, etc., are independent of the
number of employed basis functions. This means that calculations
using fixed (pre-trained) normalizing-flow parameters scale with the
truncation parameter, $P_\text{max}$ in
the same way as those using a regular linear mapping. The transferability
property significantly reduces the computational costs and makes it feasible to
apply our approach to high-dimensional systems, while maintaining accuracy
comparable to full optimization. 

\begin{figure}
 	\includegraphics[width=\columnwidth]{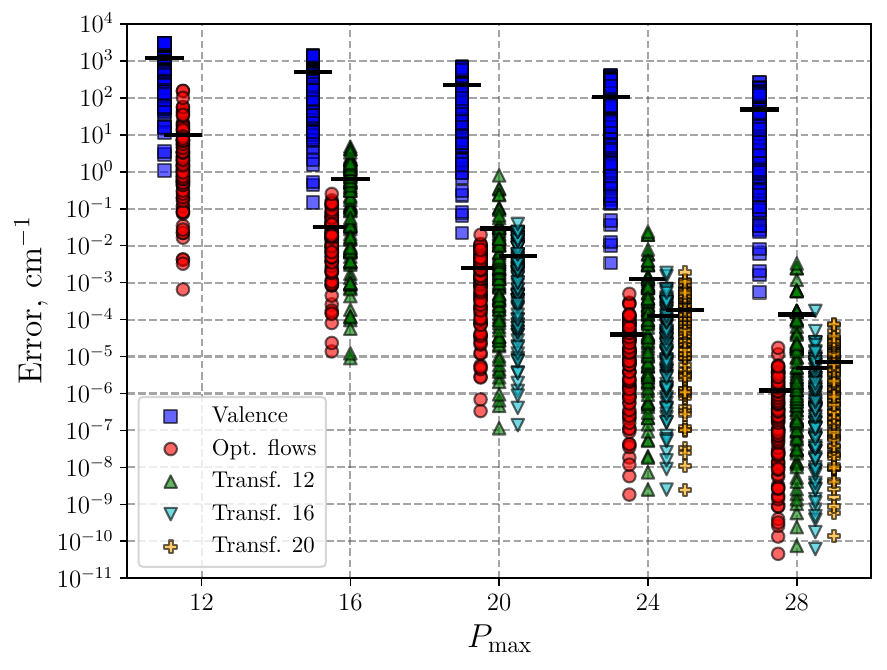}
 	\caption{\textbf{Convergence of the first 100 vibrational energies of H$_2$S.} Convergence of the first 100 energy levels of H$_2$S using valence (blue squares), and optimized normalizing-flows (red circles) coordinates, as a function of $P_\text{max}$.
    Results obtained with normalizing-flow coordinates optimized for $P_\text{max}=12$ (green up-triangles), $P_\text{max}=16$ (cyan down-triangles), and $P_\text{max}=20$ (orange plus signs) and applied to calculations with larger $P_\text{max}$ are also shown.
    Thick horizontal lines indicate the average energy-level error for each $P_\text{max}$, which is minimized through training.
    Data points are slightly offset along the $P_\text{max}$ axis for improved visual clarity.}
 	\label{fig:Transferability_H2S}
 \end{figure}

In \autoref{fig:Transferability_H2S}, we show the
convergence for the 100 first energy levels of
H$_2$S with respect to the basis-set truncation parameter, $P_\text{max}$. The results are
obtained using valence and optimized normalizing-flow coordinates.
Two types of normalizing-flow coordinates are compared: those optimized for each
specific $P_\text{max}$ (Opt. flows) and those optimized for selected values of
$P_\text{max}$ (12, 16, 20) and subsequently transferred to calculations with
larger $P_\text{max}$.
The metric for the convergence is the error of the individual energy levels
($E_i - E_i^{(\text{Ref})}$), where $E_i^{(\text{Ref})}$ represent the benchmark
energies detailed in the supplementary information. The results clearly
demonstrate that energy calculations using transferred coordinates yield greater
accuracy than those using valence coordinates. Moreover, their performance is on par with the more
computationally intensive Opt. flows coordinates. The findings also
indicate that transferring from a larger $P_{\text{max}}$ can enhance the
accuracy of highly excited states.

 The convergence of the approximation using the Hermite basis with respect to the number of basis functions, \( N \), is algebraic~\cite{Lubich:QCMD}. Specifically, the error satisfies
\[
\norm{\Psi_m - \hat{\Psi}_m} < A N^{-k},
\]
where \( \norm{\cdot} \) denotes the \( L^2 \)-norm, \( \Psi_m \) is the exact wavefunction, and \( \hat{\Psi}_m \) is its approximation. The constant \( A \) depends on the relationship between the target wavefunction and the operators associated with the Hermite basis, and \( k \) denotes the rate of convergence.
The convergence for the Hermite basis defined in normalizing-flow
coordinates is also algebraic~\cite{Saleh:thesis:2023}, with different
constants $A$ and $k$ for each map. Therefore, it is reasonable to assume that
the loss function defined in \eqref{eq:L_theta} converges algebraically, \ie,
\begin{align*}
	\mathcal{L}(N)= \frac{ \mathcal{L}_\theta^{100}(N) - \mathcal{L}_\text{Ref}^{100}}{100} \sim A \ N^{-k},
\end{align*}
where $\theta$ are the optimized parameters for the chosen $N$. To quantify the
improved convergence rate observed for the normalizing-flow coordinates (see \autoref{fig:Transferability_H2S}), we fitted $\text{log}(\mathcal{L})$ with a linear expression in $N$,
\ie, $\text{log}(\mathcal{L}) = -k \ \text{log}(N) + \text{log}(A)$. The
regression parameters derived from this fit are shown in
\autoref{tab:linear_coords}. The convergence rate of the two normalizing-flow
coordinates is significantly higher than that of the valence coordinates.
Remarkably, the convergence rate of the transferred normalizing-flow coordinates
reaches 75\% of the convergence rate of the flow coordinates optimized at each
truncation level. The constant $A$ is also
decreased by the use of nonlinear coordinates, which means that the accuracy is
improved for any fixed truncation.

\begin{table}[h!]
	\vskip 0.15in
	\centering
	\small \vskip 5pt
	\begin{tabular*}{\linewidth}{@{\extracolsep{\fill}}lcccc}
		\toprule
		Coordinate & $k \times 10^3$ & $\log(A)$  \\
		\midrule
		Valence  &  0.61$\pm$0.08 &  11.7$\pm$0.3 \\
		Opt. flows   &  2.90$\pm$0.47 &  6.43$\pm$2   \\
		Transf. 12  &  2.17$\pm$0.30 &  7.11$\pm$1   \\
		\bottomrule
	\end{tabular*}
	\caption{Convergence parameters for different coordinates.}
	\label{tab:linear_coords}
	\vskip -0.1in
\end{table}
 The results for $P_\text{max} = 12$ for H$_2$CO in  \autoref[c]{fig:convergence} were calculated with normalizing-flow coordinate optimized for $P_\text{max} = 9$. Additional results provided in the supplementary information further demonstrate the utility of the transferability property. In future work, we will elaborate on the transferability property across basis-set truncations and investigate the extension of the principle of transferability of normalizing-flow coordinates to different isotopologues and to molecular systems sharing similar structural motifs. This could potentially contribute to our understanding of intrinsic vibrational coordinates.

\section{Conclusion}
In summary, we introduced a general nonlinear parametrization for vibrational
coordinates of molecules using normalizing flows. By optimizing the flow
parameters through the variational principle, we significantly accelerated basis-set convergence, leading to more accurate vibrational energies. The improvement
is especially pronounced for highly excited and delocalized vibrational states.
The learned coordinates enhanced the separability of the Hamiltonian, which we
leveraged to improve assignment of approximate quantum numbers by projection onto direct
products of one-dimensional eigenfunctions. The enhanced separability also
potentially allows for a more intuitive interpretation of the key motifs in
strongly-coupled vibrational dynamics. The transferability of the optimized
coordinates across different truncation levels provides a computationally
efficient protocol for larger molecular system calculations. As other variational approaches, our method suffers from an exponential growth in the size of the product basis as the number of coordinates increases.
This challenge has been effectively addressed in the literature using
prescreening techniques that selectively retain only the most relevant basis-product configurations for the states of interest~\cite{Schroeder:VCIT:2021,
Billous:PRL131:133002}. It should be possible to combine the present normalizing-flow approach with state-specific eigenvalue solvers, where the basis-product configurations are tailored to specific vibrational states.
One promising method specifically designed for vibrational solutions is the iterative residuum-based RACE algorithm~\cite{Petrenko:JCP146:12}.
 After the release
of the first arXiv version of this work~\cite{Saleh:arXiv2308.16468}, another
group integrated the concept of normalizing flows for basis-set augmentation
with Monte-Carlo methods and successfully applied it to high-dimensional
systems~\cite{Zhang:JCP161:024103}.

We also explored the applicability of the normalizing-flow method for excited
electronic states, testing it on single-electron systems such as the hydrogen
atom, hydrogen molecular ion, and carbon atom in the single-active electron
approximation. Results presented in the supplementary information show a
significant improvement in basis-set convergence. This suggests promising
potential of normalizing-flow method for electronic structure problems,
especially since neural-network-based methods for excited state computations
remain challenging~\cite{Cuzzocrea:JCTC16:4203, Entwistle:NatComm14:247}.

% The present method is robust and can be readily applied for large-scale
% rovibrational spectra computations. Shortly after the first arXiv version of
% this work~\cite{Saleh:arXiv2308.16468}, another group integrated the
% normalizing-flows idea with Monte-Carlo method and successfully applied to
% high-dimensional systems~\cite{Zhang:JCP161:024103}.

% We believe the approach of combining basis sets with invertible bi-Lipschitz
% mappings could be broadly applicable, particularly in electronic structure
% correlated and density-functional theories, where neural-network-based methods
% for excited state computations remain challenging~\cite{Cuzzocrea:JCTC16:4203,
% Entwistle:NatComm14:247}.

\section*{Supporting Information}
The supporting information provides a detailed mathematical description of the
normalizing-flow approach, along with additional details on our computational
setup, reference calculations, and further investigations. It also includes an
application of the normalizing-flow method to excited electronic states.

\section*{Data availability}
The results reported in this manuscript did not depend on any specific data.

\section*{Code availability}

The code developed in this work is publicly available at \url{https://gitlab.desy.de/CMI/CMI-public/flows/-/releases/v0.1.0}. 

\section*{Acknowledgments}
We thank Jannik Eggers, Sebastian Nicolas Mendoza, and Vishnu Sanjay for useful
comments and discussions in early stages of this work.

This work was supported by Deutsches Elektronen-Synchtrotron DESY, a member of
the Helmholtz Association (HGF), including the Maxwell computational resource
operated at DESY, by the Data Science in Hamburg HELMHOLTZ Graduate School for
the Structure of Matter (DASHH, HIDSS-0002), and by the Deutsche
Forschungsgemeinschaft (DFG) through the cluster of excellence ``Advanced
Imaging of Matter'' (AIM, EXC~2056, ID~390715994). This work was supported by
Helmholtz AI computing resources (HAICORE) of the Helmholtz Association's
Initiative and Networking Fund through Helmholtz AI. This project received
funding from the European Union's Horizon Europe research and innovation
programme under the Marie Skłodowska-Curie program (no.~101155136).

\section*{Author contributions}
Y.S.\ and A.Y.\ conceptualized the work. Y.S., A.F.C., E.V., and A.Y.\ developed
the theory, produced the code, conducted the calculations, interpreted the
results, and wrote the manuscript. A.I.\ and J.K.\ contributed to the discussion
of results, writing, and proofreading of the manuscript.

\section*{Competing interest}
The authors declare no competing interests.
\section*{References}
\bibliography{string,cmi}%

\clearpage
\onecolumngrid

\section{Supplementary information: \\ Computing excited states of molecules using normalizing flows}%

\twocolumngrid

Learning an optimized vibrational coordinate system $\mathbf{r} \to g_\theta(\mathbf{r}) = \mathbf{q}$
and learning a new basis set $\{\phi_n (\mathbf{r}) \}_{n=0}^\infty \to \{\phi_n (\mathbf{q}) \sqrt{|1/\det \nabla_{\mathbf q} g_\theta^{-1}(\mathbf q)|} \}_{n=0}^\infty$ are two faces of the same coin, as seen, \eg, from
performing a change of variable in the matrix representation of the potential
$$
\mathbf{V}_{n'n}=\langle \gamma_{n'}|V|\gamma_{n} \rangle = \int \phi_{n'}^*(\mathbf q) V(g_\theta^{-1}(\mathbf q)) \phi_n(\mathbf q) \mathrm{d}\mathbf{q}.
$$
In the
following, we give a detailed description of the latter perspective, highlighting theoretical foundations and linking to the concept of normalizing
flows in machine learning. We also provide detailed information on the
numerical simulations conducted in this study, along with additional results
that can faciliate the application of this method in other domains.
\section{Mathematical foundations}

Composing a basis set $\{\phi_n (\mathbf{r}) \}_{n=0}^\infty$ with an invertible
mapping $g$ yields a sequence of functions $\{\phi_n (\mathbf{q}) \sqrt{|1/\det \nabla_{\mathbf q} g^{-1}(\mathbf q)|} \}_{n=0}^\infty$.
Certain choices of the mapping $g$ can produce sequences of functions with
improved approximation properties~\cite{Gottlieb:SpectralMethods:1977}. A
common example is the linear map $g(x) = a \ x + b$, where $a$ and $b$ are chosen to align the
potential of the problem with the potential that generates the basis set around the
equilibrium geometry of the molecule. There are also examples of nonlinear fixed
mappings, such as trigonometric or exponential functions.

However, the use of adaptive nonlinear mappings introduces additional complexity.
To ensure convergence of the method, the basis set must retain its completeness
after compositing with the mapping $g$.
The concept of composing basis sets with an invertible neural network was first introduced in
\citet{Cranmer:arXiv1904:05903}.  However, the approach was only applied to simple models, and the rationale for employing invertible neural networks was
not fully justified or rigorously explored. 

The rationale for using invertible neural networks
from an approximation-theory perspective
was first analyzed in \citet{Saleh:PAMM23:e202200239} and later rigorously
established in \citet{Saleh:arXiv2406:18613}, where sufficient and necessary
conditions for completeness under perturbation by composition operators were characterized.

Normalizing flows are foundational models in the field of generative machine
learning~\cite{Papamakarios:JMLR22:1}, rooted in
the probability integral transform initially introduced in ~\citet{Fisher:Statistical:1925}. This theorem states that any continuous probability density $p$ can be converted by a change of variables to a uniform distribution on the interval $[0,1]$. By imposing a few reasonable conditions to $p$, \eg, finiteness, it can be shown that this change of variables is bijective. As a result, any pair of probability densities $p$ and $q$ that satisfy these  conditions can be related through a change of variables, as both can be bijectively mapped into the uniform distribution.

\begin{figure}
	\centering
	\includegraphics[width=0.8\linewidth]{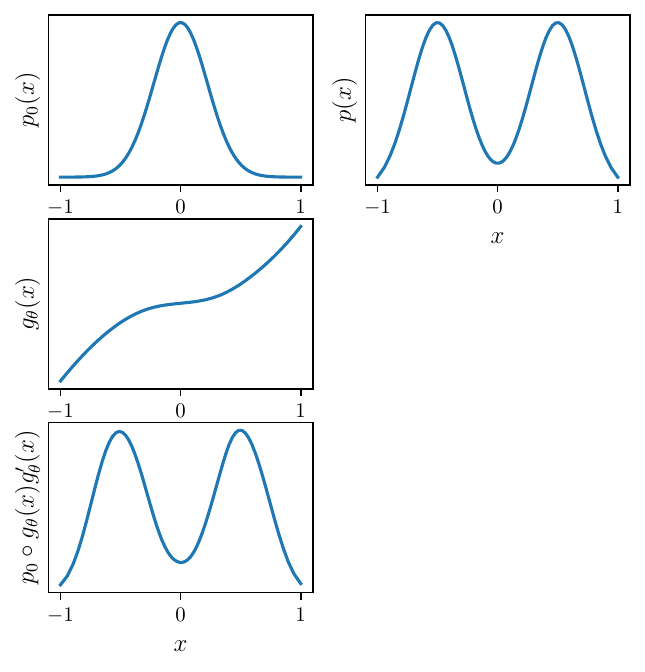}
	\caption{An illustration of normalizing flows augmenting a Gaussian distribution $p_0$.  Approximating a bimodal distribution $p$ with a single Gaussian is not possible. However, composing $p_0$ with the function $g_\theta$ produces a good approximation.}
	\label{fig:basedist}
\end{figure}

To illustrate this concept, consider the following example, which is commonly used for its simplicity. Let $p$ be an unknown bimodal probability distribution
and consider approximating it by the Gaussian
distribution $p_0$. The Gaussian distribution is a simple distribution characterized by
only two parameters, the mean and the variance. However, this approximation of $p$
\emph{via} $p_0$ is inherently flawed since no combination of the mean and
the variance can make the Gaussian $p_0$ exhibit a bimodal behavior. However, as discussed earlier, we can introduce an invertible change of variables $q = g(x)$ which allows
us to transform the Gaussian distribution $p_0$ into a more complex distribution $p^A$ that can approximate $p$. 
Specifically, we define the new distribution $p^A$ as
\begin{align*}
	p^A(x) = p_0(g(x)) |\det \nabla_x g(x)|.
\end{align*}
Multiplying by $|\det \nabla_x g(x)|$, the determinant of the Jacobian of the change of variables, ensures that the distribution remains normalized. Given that both the Gaussian and bimodal distributions are well-behaved, there exists a suitable function $g$ that can bridge this gap.
In this case, we parametrize $g$ as an invertible function $g_\theta$, thus augmenting the expressivity of the simple Gaussian
model. The results can be found in \autoref{fig:basedist}.

If the expressivity of a single Gaussian distribution can be enhanced, why
not extend this approach to a family of functions, such as a basis set?
This concept forms a core idea of our  manuscript. 

\section{Computational details}
% We provide extensive details to the simulations performed in our letter.

A summary of the atomic masses used in this study are provided in \autoref{tab:masses}.

\begin{table}[h]
    \centering
    \caption{The atomic masses utilized in calculations of the vibrational energies for the H$_2$S, H$_2$CO and HCN/HNC molecules.}
    \label{tab:masses}
    \small
    \vskip 5pt
    \begin{tabular*}{\linewidth}{@{\extracolsep{\fill}}lc}
    \toprule
    Atom & Mass (u)   \\
    %\midrule
    \hline
	Hydrogen    & 1.00782505  \\
    Sulfur & 31.97207070 \\
    Carbon    & 12.0  \\
    Oxygen & 15.99491463 \\
	Nitrogen & 14.003074004251 \\
    %\bottomrule
\hline    
\end{tabular*}
\end{table}

\subsection{Architecture of the invertible residual network}

In this work, we model the normalizing flow $g_\theta$ as an
invertible residual neural network (iResNet)~\cite{Behrmann:ICML2019:573}. An iResNet is constructed by
concatenating blocks of the form
$$
\mathbf{x}_{k+1} = \mathbf{x}_k + \mathbf{f}_k(\mathbf{x}_k; \theta),
$$
where $ \mathbf{f}_k$ is the $k$-th block, $\mathbf{x}_k$ is
the input to this block, and $\mathbf{x}_{k+1}$ is the output. We denote the initial set of
coordinates as $\mathbf{x}_0$, and for an iResNet with $K$ blocks, we
denote its output as $\mathbf{q} = \mathbf{x}_{K-1}$. Each residual
block is a standard feed-forward neural network composed of
weights, biases and nonlinear activations. Each such block is guaranteed
to be invertible if $ \mathbf{f}_k$ is Lipschitz, with Lipschitz constant $< 1$. The
inverse of each block can be computed by a fixed-point iteration method. 

To guarantee that the blocks $ \mathbf{f}_k$ for $k=0, \dots, K-1$
satisfy the Lipschitz condition, we used the LipSwish activation
function
$$
\sigma(x) := \left(\frac{1}{1.1}\right) \frac{1}{1+\exp(-x)},
$$
which has a Lipschitz constant $\approx 1$. A block $\mathbf{f}_k$ containing such
activation functions is thus guaranteed to be Lipschitz if each of the linear passes has a Lipschitz constant $<1$, \ie, the weight
matrices $W$ are normalized to have a spectral norm $<1$. This is achieved by setting 
\begin {align*}
W = \begin{cases}
	W &\quad\text{if } \|W\|_2 < c\\
	c \cdot \frac{W}{\|W\|_2} &\quad\text{if } \|W\|_2 \geq c,
\end{cases}
\end{align*}
where $0<c<1$ is a hyperparameter and $\|W\|_2$ is the
spectral norm of the matrix $W$. For a block $ \mathbf{f}_k$ containing $n$ hidden layers,
it can be shown that
\begin{align*}
\mathrm{Lip}( \mathbf{f}_k) 
&\leq c^{n+1}.
\end{align*}

A special attention must be given to the architecture of each block. While a block with a Lipschitz constant close to 1 allows for
a greater flexibility, a higher Lipschitz constant can significantly hinder the convergence of the
fixed-point iteration method used for computing its inverse. To balance
these effects, we used blocks with $2$ hidden layers for all of our calculations
and set $c=0.9$, producing blocks with Lipschitz constants $\sim0.7$.

\subsection{Comparison against reference data}\label{sec:comparison}

Benchmark
results for vibrational energy calculations were generated using the wavefunction ansatz described in
 Eq.~3 of the manuscript. The calculations employed a basis polyad truncation $P_{\text{max}}=60$ for H$_2$S,
$P_{\text{max}}=16$ for H$_2$CO, and $P_{\text{max}}=44$ and for HCN/HCN.
We compared our results with those of ~\citet{Azzam:MNRAS460:4063}
for H$_2$S, \citet{Al-Refaie:MNRAS448:1704} for H$_2$CO, and
\citet{Mourik:JCP115:3706} for HCN/HNC.

In \autoref{tab:1}, we present the calculated energies $E_i$, the reference values $E_i^{\text{ref}}$ and the deviations $\Delta E_i = E_i-E_i^{\text{ref}}$
across the $i=1..100$ lowest energy levels for the three molecular systems. The results demonstrate almost
exact agreement with the reference data, with most of the states converging to even lower limits than
the reference data.

\LTcapwidth=\textwidth
\begin{longtable*}{@{\extracolsep{\fill}}cccccccccc}
\caption{Comparison of vibrational energies (in \invcm) of H$_2$S, HCN/HNC, and H$_2$CO molecules,
	calculated using the normalizing-flow approach $E_i$, with reference energies $E_i^{\text{ref}}$~\cite{Azzam:MNRAS460:4063, Mourik:JCP115:3706, Al-Refaie:MNRAS448:1704}, across the 100 lowest vibrational states. A positive deviation $\Delta E_i$ indicates a better convergence is achieved by our prediction.}
\label{tab:1}
	\endfirsthead
	\endhead
	\hline
	\multirow{2}{*}{{\centering $i$ }} & 
	\multicolumn{3}{c|}{\centering \textbf{H\textsubscript{2}S}} & 
	\multicolumn{3}{c|}{\textbf{HCN}} & 
	\multicolumn{3}{c}{\textbf{H\textsubscript{2}CO}} \\
	\cline{2-10}
	& $E_i$ & $E_i^{\text{ref}}$ & 
	$\Delta E_i$ & $E_i$ & $E_i^{\text{ref}}$ & 
	$\Delta E_i$ & 
	$E_i$ & $E_i^{\text{ref}}$ & 
	$\Delta E_i$  \\
	\hline
	1 & 0.00 & 0.00 & 0.00 & 0.00 & 0.00 & 0.00 & 0.00 & 0.00 & 0.00 \\
	2 & 1182.57 & 1182.58 & -0.01 & 1414.92 & 1414.92 & -0.00 & 1167.29 & 1167.34 & -0.05 \\
	3 & 2353.91 & 2353.96 & -0.06 & 2100.58 & 2100.58 & 0.00 & 1249.06 & 1249.07 & -0.01 \\
	4 & 2614.39 & 2614.41 & -0.01 & 2801.46 & 2801.46 & -0.00 & 1500.10 & 1500.12 & -0.03 \\
	5 & 2628.46 & 2628.45 & 0.01 & 3307.74 & 3307.75 & -0.01 & 1746.03 & 1746.05 & -0.02 \\
	6 & 3513.70 & 3513.79 & -0.09 & 3510.99 & 3510.99 & 0.00 & 2327.34 & 2327.50 & -0.15 \\
	7 & 3779.19 & 3779.17 & 0.02 & 4176.24 & 4176.24 & 0.00 & 2422.56 & 2422.63 & -0.07 \\
	8 & 3789.27 & 3789.27 & 0.00 & 4181.46 & 4181.45 & 0.01 & 2494.25 & 2494.32 & -0.07 \\
	9 & 4661.61 & 4661.67 & -0.07 & 4686.28 & 4686.29 & -0.01 & 2666.81 & 2667.04 & -0.22 \\
	10 & 4932.69 & 4932.70 & -0.01 & 4891.76 & 4891.76 & 0.00 & 2718.97 & 2719.08 & -0.12 \\
	11 & 4939.13 & 4939.10 & 0.03 & 5185.57 & 5185.64 & -0.07 & 2782.37 & 2782.41 & -0.04 \\
	12 & 5145.03 & 5144.99 & 0.05 & 5394.43 & 5394.43 & 0.00 & 2843.32 & 2843.34 & -0.01 \\
	13 & 5147.17 & 5147.22 & -0.05 & 5537.76 & 5537.76 & -0.00 & 2905.75 & 2905.86 & -0.11 \\
	14 & 5243.16 & 5243.10 & 0.06 & 5586.50 & 5586.50 & 0.00 & 2998.91 & 2999.01 & -0.10 \\
	15 & 5797.21 & 5797.23 & -0.03 & 6033.72 & 6033.72 & -0.00 & 2999.96 & 3000.00 & -0.04 \\
	16 & 6074.57 & 6074.58 & -0.02 & 6127.51 & 6127.56 & -0.05 & 3238.84 & 3238.94 & -0.10 \\
	17 & 6077.63 & 6077.59 & 0.03 & 6242.43 & 6242.42 & 0.00 & 3471.66 & 3471.72 & -0.06 \\
	18 & 6288.13 & 6288.15 & -0.01 & 6260.59 & 6260.59 & -0.00 & 3480.98 & 3481.30 & -0.31 \\
	19 & 6289.13 & 6289.17 & -0.04 & 6513.48 & 6513.50 & -0.02 & 3585.63 & 3585.89 & -0.26 \\
	20 & 6385.32 & 6385.32 & -0.00 & 6768.51 & 6768.51 & -0.00 & 3675.03 & 3675.21 & -0.18 \\
	21 & 6920.08 & 6920.08 & -0.00 & 6879.60 & 6879.60 & 0.00 & 3736.63 & 3737.05 & -0.42 \\
	22 & 7204.31 & 7204.31 & -0.00 & 6960.99 & 6960.99 & -0.00 & 3825.37 & 3825.97 & -0.60 \\
	23 & 7204.43 & 7204.44 & -0.00 & 7088.71 & 7088.74 & -0.03 & 3887.05 & 3887.36 & -0.31 \\
	24 & 7419.85 & 7419.85 & -0.00 & 7210.51 & 7210.59 & -0.08 & 3935.85 & 3936.44 & -0.59 \\
	25 & 7420.08 & 7420.09 & -0.02 & 7369.18 & 7369.18 & -0.00 & 3941.30 & 3941.53 & -0.23 \\
	26 & 7516.83 & 7516.83 & -0.00 & 7461.59 & 7461.60 & -0.01 & 3996.26 & 3996.48 & -0.23 \\
	27 & 7576.41 & 7576.38 & 0.03 & 7617.24 & 7617.24 & 0.00 & 4022.46 & 4022.56 & -0.09 \\
	28 & 7576.60 & 7576.54 & 0.05 & 7641.28 & 7641.28 & 0.00 & 4057.79 & 4058.10 & -0.31 \\
	29 & 7752.34 & 7752.26 & 0.08 & 7855.83 & 7855.84 & -0.01 & 4083.44 & 4083.49 & -0.05 \\
	30 & 7779.35 & 7779.32 & 0.03 & 8020.41 & 8020.45 & -0.04 & 4163.89 & 4164.13 & -0.24 \\
	31 & 8029.81 & 8029.81 & -0.00 & 8110.25 & 8110.25 & -0.00 & 4164.61 & 4165.27 & -0.66 \\
	32 & 8318.68 & 8318.68 & -0.00 & 8140.64 & 8140.69 & -0.05 & 4192.77 & 4193.19 & -0.41 \\
	33 & 8321.86 & 8321.86 & -0.00 & 8197.55 & 8197.55 & -0.00 & 4247.40 & 4247.61 & -0.21 \\
	34 & 8539.57 & 8539.56 & 0.00 & 8283.37 & 8283.37 & 0.00 & 4256.12 & 4256.31 & -0.19 \\
	35 & 8539.82 & 8539.82 & -0.00 & 8323.52 & 8323.52 & -0.00 & 4335.01 & 4335.11 & -0.10 \\
	36 & 8637.16 & 8637.16 & -0.00 & 8584.71 & 8584.74 & -0.03 & 4397.64 & 4398.13 & -0.49 \\
	37 & 8697.13 & 8697.14 & -0.01 & 8691.66 & 8691.67 & -0.01 & 4466.76 & 4467.12 & -0.35 \\
	38 & 8697.18 & 8697.16 & 0.02 & 8830.27 & 8830.27 & -0.00 & 4495.16 & 4495.50 & -0.34 \\
	39 & 8878.59 & 8878.59 & -0.00 & 8850.65 & 8850.74 & -0.09 & 4529.49 & 4529.64 & -0.15 \\
	40 & 8897.38 & 8897.38 & -0.00 & 8945.48 & 8945.50 & -0.02 & 4571.53 & 4571.66 & -0.12 \\
	41 & 9126.09 & 9126.09 & -0.00 & 8954.46 & 8954.47 & -0.01 & 4623.65 & 4623.93 & -0.28 \\
	42 & 9420.24 & 9420.24 & -0.00 & 9009.00 & 9009.00 & 0.00 & 4628.67 & 4629.22 & -0.55 \\
	43 & 9426.39 & 9426.39 & -0.00 & 9088.01 & 9088.05 & -0.04 & 4729.67 & 4730.04 & -0.37 \\
	44 & 9647.10 & 9647.17 & -0.07 & 9164.06 & 9164.08 & -0.02 & 4734.14 & 4734.33 & -0.19 \\
	45 & 9647.61 & 9647.61 & -0.00 & 9214.77 & 9214.85 & -0.08 & 4740.89 & 4741.40 & -0.52 \\
	46 & 9745.80 & 9745.80 & -0.00 & 9440.11 & 9440.12 & -0.01 & 4840.27 & 4840.80 & -0.53 \\
	47 & 9806.71 & 9806.67 & 0.05 & 9488.55 & 9488.54 & 0.01 & 4926.09 & 4926.53 & -0.44 \\
	48 & 9806.75 & 9806.73 & 0.01 & 9508.91 & 9508.91 & -0.00 & 4956.57 & 4956.91 & -0.34 \\
	49 & 9911.10 & 9911.02 & 0.08 & 9619.20 & 9619.29 & -0.09 & 4977.12 & 4978.25 & -1.12 \\
	50 & 9911.11 & 9911.02 & 0.09 & 9674.67 & 9674.67 & -0.00 & 4977.37 & 4978.44 & -1.08 \\
	51 & 9993.68 & 9993.68 & -0.00 & 9675.25 & 9675.25 & 0.00 & 5041.43 & 5042.22 & -0.78 \\
	52 & 10004.98 & 10004.98 & -0.00 & 9743.71 & 9743.79 & -0.08 & 5092.00 & 5092.58 & -0.58 \\
	53 & 10188.36 & 10188.36 & -0.00 & 9865.30 & 9865.33 & -0.03 & 5108.70 & 5109.39 & -0.68 \\
	54 & 10194.51 & 10194.45 & 0.06 & 9922.90 & 9922.92 & -0.02 & 5140.25 & 5141.01 & -0.76 \\
	55 & 10208.77 & 10208.77 & -0.00 & 9993.94 & 9993.95 & -0.01 & 5153.10 & 5154.39 & -1.29 \\
	56 & 10292.54 & 10292.54 & -0.00 & 10006.55 & 10006.59 & -0.04 & 5177.42 & 5177.65 & -0.24 \\
	57 & 10508.45 & 10508.45 & -0.00 & 10132.33 & 10132.38 & -0.05 & 5186.87 & 5187.25 & -0.38 \\
	58 & 10517.59 & 10517.59 & -0.00 & 10165.62 & 10165.60 & 0.02 & 5204.40 & 5205.00 & -0.61 \\
	59 & 10742.14 & 10742.14 & -0.00 & 10266.38 & 10266.40 & -0.02 & 5246.10 & 5246.39 & -0.29 \\
	60 & 10742.73 & 10742.73 & -0.00 & 10304.11 & 10304.10 & 0.01 & 5255.85 & 5256.28 & -0.44 \\
	61 & 10842.18 & 10842.18 & -0.00 & 10364.93 & 10364.90 & 0.03 & 5312.30 & 5312.86 & -0.57 \\
	62 & 10904.69 & 10904.69 & -0.00 & 10460.51 & 10460.50 & 0.01 & 5320.86 & 5322.42 & -1.56 \\
	63 & 10904.77 & 10904.77 & -0.00 & 10636.68 & 10636.70 & -0.02 & 5324.95 & 5325.30 & -0.35 \\
	64 & 11008.77 & 11008.70 & 0.08 & 10654.81 & 10654.87 & -0.06 & 5357.25 & 5358.09 & -0.84 \\
	65 & 11008.79 & 11008.77 & 0.01 & 10749.73 & 10749.70 & 0.03 & 5385.38 & 5386.43 & -1.05 \\
	66 & 11097.17 & 11097.17 & -0.00 & 10753.95 & 10753.98 & -0.03 & 5415.13 & 5415.83 & -0.70 \\
	67 & 11101.53 & 11101.53 & -0.00 & 10758.04 & 10758.00 & 0.04 & 5417.34 & 5417.91 & -0.56 \\
	68 & 11278.08 & 11278.08 & -0.00 & 10861.98 & 10862.15 & -0.17 & 5432.57 & 5433.31 & -0.75 \\
	69 & 11291.96 & 11291.96 & -0.00 & 10871.09 & 10871.10 & -0.01 & 5462.80 & 5463.15 & -0.35 \\
	70 & 11294.93 & 11294.93 & -0.00 & 10922.00 & 10922.03 & -0.03 & 5489.37 & 5490.04 & -0.67 \\
	71 & 11391.68 & 11391.68 & -0.00 & 10925.18 & 10925.30 & -0.12 & 5489.92 & 5490.58 & -0.65 \\
	72 & 11582.88 & 11582.88 & -0.00 & 11006.52 & 11006.50 & 0.02 & 5532.31 & 5532.38 & -0.06 \\
	73 & 11595.14 & 11595.14 & -0.00 & 11035.70 & 11035.70 & 0.00 & 5543.86 & 5544.61 & -0.75 \\
	74 & 11824.13 & 11824.13 & -0.00 & 11065.18 & 11065.23 & -0.05 & 5552.06 & 5552.88 & -0.82 \\
	75 & 11824.65 & 11824.65 & -0.00 & 11198.41 & 11198.51 & -0.10 & 5625.66 & 5626.49 & -0.83 \\
	76 & 11925.74 & 11925.75 & -0.00 & 11225.82 & 11225.90 & -0.08 & 5651.04 & 5651.16 & -0.13 \\
	77 & 11990.55 & 11990.55 & -0.00 & 11271.68 & 11271.70 & -0.02 & 5659.48 & 5660.99 & -1.51 \\
	78 & 11990.65 & 11990.65 & -0.00 & 11489.39 & 11489.40 & -0.01 & 5666.66 & 5667.46 & -0.80 \\
	79 & 12095.40 & 12095.40 & -0.00 & 11536.09 & 11536.10 & -0.01 & 5680.97 & 5681.46 & -0.50 \\
	80 & 12095.44 & 12095.44 & -0.00 & 11539.63 & 11539.71 & -0.08 & 5687.28 & 5688.23 & -0.96 \\
	81 & 12149.52 & 12149.46 & 0.06 & 11549.66 & 11549.60 & 0.06 & 5718.43 & 5718.83 & -0.40 \\
	82 & 12149.55 & 12149.52 & 0.04 & 11589.99 & 11590.02 & -0.03 & 5731.64 & 5732.19 & -0.55 \\
	83 & 12186.40 & 12186.40 & -0.00 & 11672.67 & 11672.80 & -0.13 & 5766.47 & 5766.91 & -0.44 \\
	84 & 12188.55 & 12188.55 & -0.00 & 11688.05 & 11688.10 & -0.05 & 5768.41 & 5769.06 & -0.65 \\
	85 & 12334.64 & 12334.64 & -0.00 & 11710.13 & 11710.10 & 0.03 & 5770.76 & 5771.67 & -0.92 \\
	86 & 12383.69 & 12383.69 & -0.00 & 11744.32 & 11744.36 & -0.04 & 5809.52 & 5809.85 & -0.33 \\
	87 & 12384.60 & 12384.60 & -0.00 & 11744.36 & 11744.50 & -0.14 & 5822.38 & 5822.71 & -0.33 \\
	88 & 12481.05 & 12481.05 & -0.00 & 11832.45 & 11832.49 & -0.04 & 5887.13 & 5888.44 & -1.31 \\
	89 & 12524.83 & 12524.64 & 0.20 & 11969.83 & 11969.90 & -0.07 & 5887.96 & 5888.55 & -0.59 \\
	90 & 12525.35 & 12525.21 & 0.13 & 11970.23 & 11970.27 & -0.04 & 5889.15 & 5890.16 & -1.01 \\
	91 & 12643.27 & 12643.27 & -0.00 & 11977.84 & 11977.70 & 0.14 & 5936.19 & 5937.06 & -0.87 \\
	92 & 12658.92 & 12658.92 & -0.00 & 12055.72 & 12055.70 & 0.02 & 5984.87 & 5985.33 & -0.46 \\
	93 & 12695.20 & 12695.20 & -0.00 & 12102.62 & 12102.69 & -0.07 & 5987.32 & 5988.05 & -0.73 \\
	94 & 12735.21 & 12735.21 & 0.00 & 12192.45 & 12192.50 & -0.05 & 5996.72 & 5997.93 & -1.21 \\
	95 & 12892.52 & 12892.52 & -0.00 & 12200.53 & 12200.60 & -0.07 & 5998.74 & 5999.46 & -0.72 \\
	96 & 12892.83 & 12892.83 & -0.00 & 12304.49 & 12304.50 & -0.01 & 6053.12 & 6053.56 & -0.44 \\
	97 & 12995.95 & 12995.95 & -0.00 & 12311.73 & 12311.70 & 0.03 & 6091.73 & 6093.56 & -1.83 \\
	98 & 13063.69 & 13063.69 & -0.00 & 12374.87 & 12375.11 & -0.24 & 6107.28 & 6108.32 & -1.04 \\
	99 & 13063.80 & 13063.80 & -0.00 & 12382.04 & 12382.00 & 0.04 & 6122.55 & 6124.52 & -1.96 \\
	100 & 13170.37 & 13170.37 & -0.00 & 12384.74 & 12384.75 & -0.01 & 6177.20 & 6178.84 & -1.63 \\
	\hline
\end{longtable*}

% \twocolumngrid

\section{Further Investigations}
\subsection{Convergence with respect to enhancement of normalizing-flow complexity}
\begin{figure}
	\includegraphics[width=\linewidth]{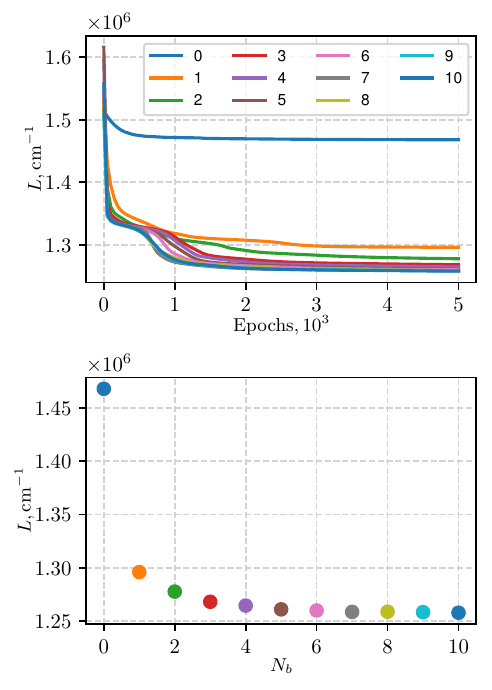}%
	\caption{Convergence of the sum of the lowest 100 vibrational energy levels $L=\sum_i^{100}E_i$ for HCN/HNC as a function of the number of blocks in the iResNet model $N_b$ using a basis set truncated at $P_{\text{max}}=16$. The upper plot shows the convergence of this sum over the number of optimization epochs, with results for different $N_b$ represented by different colors. The lower plot shows the convergence of the optimized sum as a function of $N_b$.}
	\label{fig:complexity}
\end{figure}

In spectral methods, the standard approach for improving the accuracy of computed energy levels is to increase the number of basis functions $N$. However,
this can be computationally expensive, with memory costs scaling as $N^2$ and certain computational tasks, such as diagonalization of the Hamiltonian matrix, scaling up to $N^3$.
The normalizing-flow approach offers a possibility of enhancing the expressivity of the basis functions by creating a more complex mapping, achieved by adding additional blocks. In \autoref{fig:complexity}, we show the sum of the 100 lowest vibrational energies (loss) of the HCN/HNC isomers as a function of the number of blocks in the iResNet model, for a fixed number of basis functions (at $P_{\text{max}}=16$). 

\subsection{Sensitivity of normalizing-flow coordinates to the number of target states}
\begin{figure}
	\includegraphics[width=\linewidth]{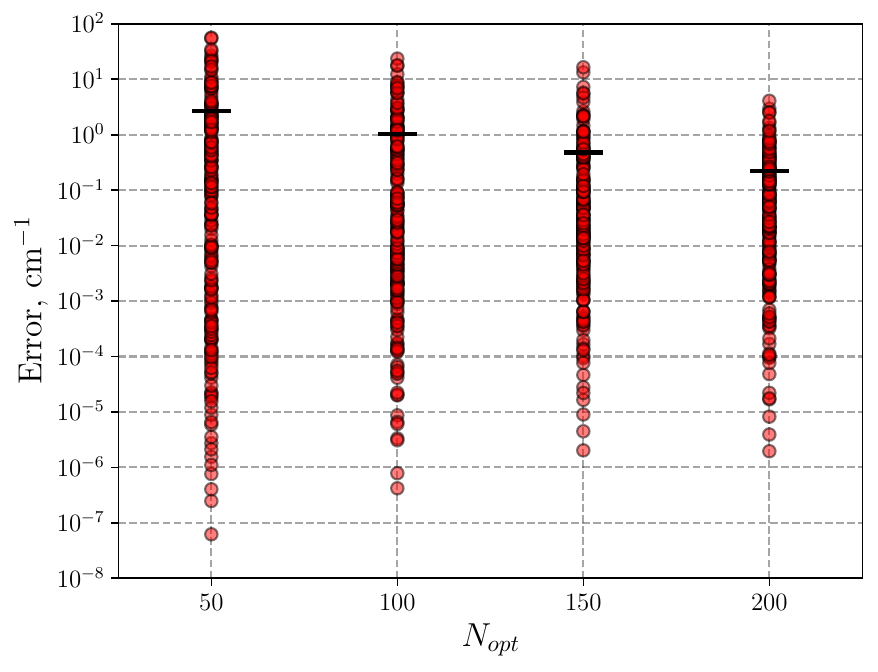}%
	\caption{Convergence of the lowest 200 vibrational energy levels $(E_i,
		i=1..200)$ for H$_2$S, calculated using $P_{\text{max}}=20$ (506 basis functions) and normalizing-flow coordinates. The energy discrepancies $(\Delta E_i)$ relative to our converged
		benchmark reference are shown for normalizing-flow coordinates optimized for different numbers of target states $(N_{opt})$. The solid black lines show the average energy discrepancy.}
	\label{fig:diff_target_states}
\end{figure}

In this subsection, we examine the impact of optimizing normalizing-flow
coordinates for different numbers of target states. In
\autoref{fig:diff_target_states}, we show the convergence of the first 200
vibrational energy levels of H$_2$S, computed with normalizing-flow coordinates
with $P_{max}=20$, corresponding to 506 basis functions. The coordinates were
optimized to minimize the lowest 50, 100, 150, and 200 target states, respectively. As shown in
\autoref{fig:diff_target_states}, the average error of the first 200 energy
levels decreases as the number of target states increases. The decrease in
average energy error is accomplished by adjusting the coordinates to better
converge the highest energy states in their respective subsets. However, as the
number of target states increases, this refinement slightly compromises the
convergence of the lowest energy states, which contribute progressively less to
the loss function.

\subsection{Transferability of the normalizing flow across different basis set truncation levels}
\begin{figure}
    \includegraphics[width=\linewidth]{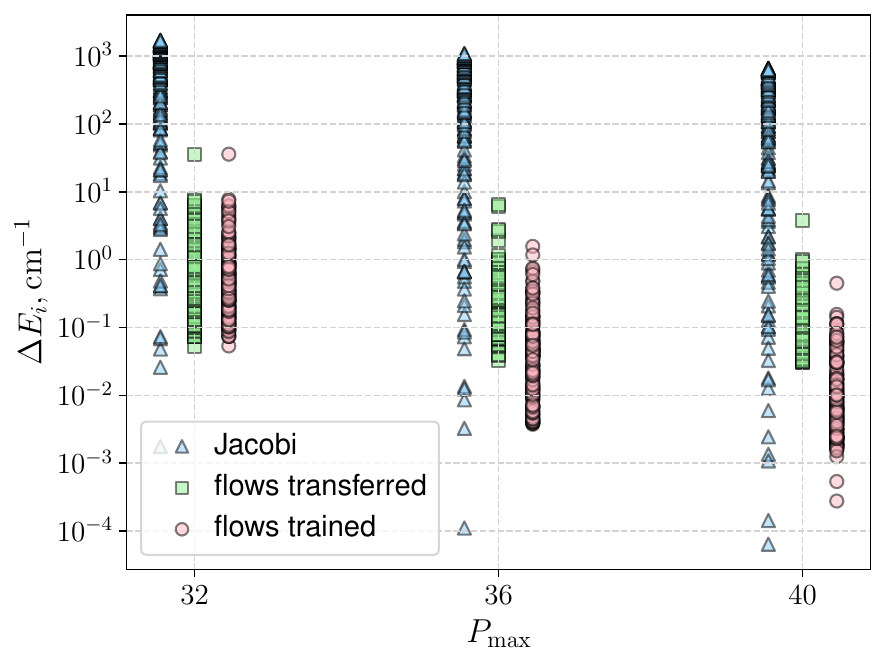}%
    \caption{Convergence of the lowest 200 vibrational energy levels $(E_i,
    i=1..200)$ for HCN/HNC, calculated using Jacobi coordinates (triangles,
    light blue),
    normalizing-flow coordinates transferred from $P_{\text{max}}=32$ (squares, light
    green), and optimized normalizing-flow coordinates (circles, light
    pink). The energy discrepancies $(\Delta E_i)$ relative to our converged
    benchmark reference are shown for basis sets truncated at
    $P_{\text{max}}=32$ (1785 basis functions), 36 (2470), and 40 (3311). All states
    are slightly offset along the $P_\text{max}$ axis for visual clarity.}
    \label{fig:transfer}
\end{figure}

During our investigations into the interpretability of the normalizing-flow
coordinates, we noticed that the optimized coordinates remain consistent across different
basis set truncations, parametrized by $P_{\text{max}}$. This observation prompted us to explore
the transferability of the learned mapping across different truncation levels.
Interestingly, we found that a normalizing flow trained with few basis functions could effectively transfer to
larger basis sets
without the need for retraining. In \autoref{fig:transfer}, we demonstrate this by
comparing the vibrational energy levels of HCN/HNC using a normalizing flow initially
trained to minimize the sum of the lowest 200 energies with a basis truncated at $P_{\text{max}}=32$. This mapping was then used with basis sets truncated at $P_\text{max}>32$ and compared to the results of
calculations where the normalizing flow was optimized for each truncation level.

The results in \autoref{fig:transfer} reveal that the transferred normalizing-flow mapping provides comparable accuracy to the optimized mapping for each truncation level and performs significantly better than Jacobi coordinates. 

It is worth highlighting that transferability enables significant computational savings. Expanding calculations to incorporate a larger number of basis functions is feasible as long as the quantities dependent on normalizing flow, such as $\frac{\partial q_{\alpha}}{\partial r_l}$, $\frac{\partial D}{\partial r_l}$, $\frac{1}{D}$, etc, can be efficiently stored in memory or recomputed on the fly. Importantly, the size and computation cost of these quantities are independent of the basis size, meaning that
calculations using a pre-trained normalizing flow scale with basis size in the same way as those using a regular linear mapping.

Based on these observations, we propose the following efficient protocol for
implementing our approach: (i) Train a normalizing flow on a small basis set and store the optimized parameters. (ii) Transfer the learned coordinates to a larger
basis set where training would be computationally prohibitive, then solve the eigenvalue problem using the pre-trained
normalizing flow to obtain accurate energy levels. This protocol is particularly useful for high dimensional systems, where
training costs are significantly higher.
The results for H$_2$S and H$_2$CO in~\autoref{tab:1} where obtained using this protocol, being transferred from the optimization at $P_{max} = 12$ and $P_{max}=9$ respectively.

\subsection{Electronic calculations}
The nonlinear ansatz described in Eq.~3 of the manuscript can be used to solve the electronic Schrödinger equation. 
To demonstrate this, we
computed electronic states of prototypical
one-electron systems such as the hydrogen atom, H$_2^+$ molecular ion, and carbon atom in the
single-active electron approximation. We only consider normalizing flows for the radial coordinate and integrated out the
angular coordinates using the spherical-harmonic basis. This approach is consistent with common
practices in quantum chemistry, where the radial basis typically presents the primary challenge and
is the main target of optimization.
As a starting basis we explored Hermite basis, the basis of
three-dimensional isotropic harmonic oscillator, and the standard atomic-orbital basis sets
from Dunning’s family~\cite{Dunning:JCP90:1007, Kendall:JCP96:6796}. The number
of basis functions was
defined by the number of radial functions. 

The Hamiltonians for the one-electron radial problems were
derived by integrating out the angular coordinates using spherical-harmonic functions $Y_{l,m}$ as the angular basis, or solid-harmonic functions $r^lY_{l,m}$ in the case of atomic-orbital basis sets.
For the Hermite and isotropic harmonic oscillator radial basis sets, we utilized a direct product of the radial and angular basis functions with $l\leq l_\text{max}$, where $l_\text{max}=3$ for atoms and 8 for molecules.
For the atomic-orbital basis set, we employed specific combinations of radial Gaussian and solid-harmonic functions as dictated by the structure of the basis sets.

The angular matrix elements of the electron-nuclei Coulomb-attraction potential were computed using
the Laplace expansion, with truncation determined by $l_\text{max}$ of the angular basis.

The electronic energies of multi-electron systems, \eg, carbon atom, were calculated using the single-active-electron
approximation, in which the electron-electron Coulomb-interaction potential is approximated by the
one-electron electrostatic potential created by the electron density $\rho(\mathbf{r})$ of a
singly-charged ion in the ground state, \ie,
\begin{align*}
V(\mathbf{r}) = -\int \frac{\rho(\mathbf{r'})}{|\mathbf{r} - \mathbf{r'}|}d\mathbf{r'}.
\end{align*}
In addition, the exchange-interaction potential is generally added using some popular approximations, like the local-density approximation models.
Our focus is primarily on investigating the basis set convergence of energies and its enhancement \emph{via} normalizing flows. The absolute accuracy of the electronic energies was not our primary concern, hence we did not include the exchange-interaction potential in our calculations.

In simulations for carbon atom, the one-electron electrostatic potential of C$^+$ was calculated using the second-order approximate coupled-cluster CC2 level of theory with the aug-cc-pV5Z atomic-orbital basis set~\cite{Dunning:JCP90:1007, Kendall:JCP96:6796}, as implemented in the Psi4 quantum chemistry package~\cite{Psi4_2020}.
The calculations were performed on a grid generated by the direct product of an equidistant radial grid $\{r_g\}_{g}^{N_g}$ and the Lebedev quadrature grid of 131st order represented by a set of angular points $\{\theta_h,\phi_h\}_{h}^{N_h}$ ~\cite{Lebedev:DoklMath:59:477}.
The radial potential was calculated by integrating angular coordinates at each radial point in the basis of spherical-harmonic functions using Lebedev quadrature rule, \ie,
\begin{align*}
&V_{l'm',lm}(r_g) = \langle Y_{l'm'}(\theta,\phi)|V(r_g,\theta,\phi)| Y_{lm}(\theta,\phi) \rangle \\ \nonumber
&\approx \sum_h^{N_h} w_h Y_{l'm'}^*(\theta_h,\phi_h)Y_{lm}(\theta_h,\phi_h)V(r_g,\theta_h,\phi_h),
\end{align*}
where $w_h$ is the Lebedev quadrature weight including spherical volume element.
For computing the angular integrals between the spherical harmonics centered at different centers, which are needed in atomic-orbital-basis calculations for molecules, we also implemented Becke's partitioning scheme~\cite{Becke:JCP88:2547}.
The calculated values of the radial potential $V_{l'm',lm}(r_g)$ were interpolated across the radial grid $\{r_g\}_{g}^{N_g}$ using the regular grid interpolator technique.

The reference energies for the hydrogen atom are known
analytically, while for the hydrogen-molecule cation H$_2^+$, the energies calculated using the
Riccati-Padé method~\cite{Fernandez:ChemSelect6:9527} were employed as the reference. In \autoref{tab:comparison_el}, we present the absolute errors for the ground and four lowest
excited electronic states of H atom and H$_2^+$, as calculated using normalizing flows with
different basis sets. For comparison, the table also includes errors corresponding to
full-configuration interaction calculations in large atomic-orbital basis sets aug-cc-pV5Z and
aug-cc-pV6Z~\cite{Dunning:JCP90:1007, Kendall:JCP96:6796}, as computed using the Psi4 quantum
chemistry package~\cite{Psi4_2020}.

\begin{table}
	\centering
	\small \vskip 5pt
	\begin{tabular*}{\linewidth}{@{\extracolsep{\fill}}lrrr}
		\toprule
		\multicolumn{4}{c}{H atom} \\
		\midrule
		State & Hermite & iso-HO & AV6Z  \\
		\midrule
		Ground   & -4.8 & -0.02 & -0.0007 \\
		1st excited   & -0.07 & -0.002 & -0.08 \\
		2d excited  & -0.07 & -0.002 & -1.22 \\
		3d excited    & -0.07 & -0.002 & -1.22 \\
		4th excited    & -0.8 & -0.002 & -1.22 \\
		\midrule
		\multicolumn{4}{c}{H$_2^+$ ion} \\
		\midrule
		State & Hermite & AV5Z  \\
		\midrule
		Ground   & -3.1 & -0.01 \\
		1st excited   & -1.7 & -0.01 \\
		2d excited  & -0.007 & -0.44 \\
		3d excited    & -0.007 & -0.44 \\
		4th excited    & -0.62 & -0.53 \\
		\bottomrule
	\end{tabular*}
	\caption{The errors ($\text{exact}-\text{calculated}$, in mHartree) in the ground and excited
	state calculations for H atom and H$_2^+$ ($R_\text{H--H}=2$~Bohr), using normalizing flows with Hermite and
	three-dimensional isotropic harmonic oscillator (iso-HO) basis sets. For comparison, errors
	corresponding to full-configuration interaction approach with aug-cc-pV6Z (AV6Z) basis for H
	and aug-cc-pV5Z (AV5Z) basis for H$_2^+$ are also listed. Note that the normalizing-flow
	calculations of H$_2^+$ employed s single-center spherical-harmonic basis truncated at
	$l_\text{max}=8$, which accounts for estimated $-1.1,-0.2,$ $-10^{-5},-10^{-5},-0.15$~mHartree
	differences in the corresponding electronic energies. }
\label{tab:comparison_el}
\end{table}

\begin{figure}
	\centering
	\includegraphics[width=\linewidth]{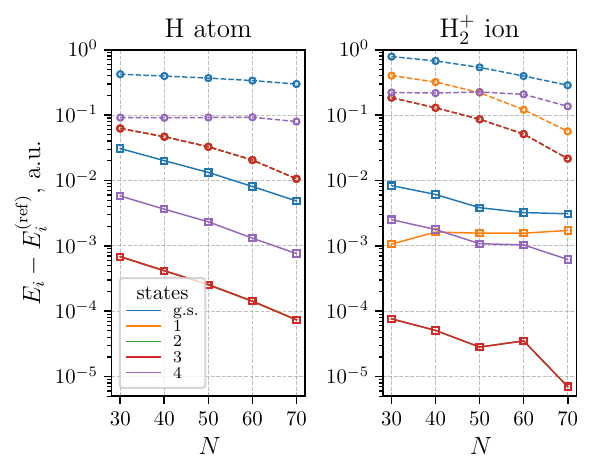}
	\caption{Convergence of the ground (g.s.) and four lowest excited electronic states of hydrogen atom and hydrogen molecular ion H$_2^+$  ($R_\text{H--H}=2$~Bohr) plotted against the number of Hermite radial basis functions $N$. Results obtained with optimized-linear (dashed, circles) and normalizing-flow (solid, squares) parameterizations are compared. The reference energies $E_i^\text{ref}$ for H$_2^+$ are from~\citet{Fernandez:ChemSelect6:9527}.
	}
	\label{fig:h_h2plus}
\end{figure}

The application of the normalizing-flow ansatz with Hermite basis set  demonstrates
high accuracy and fast basis-set convergence for the radial electronic problem, although such basis
is typically considered unsuitable for solving electronic problems. In \autoref{fig:h_h2plus}, we show the convergence of the electronic energies for H atom and H$_2^+$ molecular ion as a
function of the number of Hermite radial basis functions $N$. The results are
compared against the linear parametrization, obtained by setting $g({\mathbf x})
= \mathbf{a}\cdot{\mathbf x}+\mathbf{b}$ in Eq.~3 of the manuscript, where the linear parameters $\mathbf a$ and $\mathbf b$
were optimized. Using the normalizing-flow ansatz, the
electronic energies of H and H$_2^+$ converge very quickly to within few mHartree of the exact
values for the ground state, respectively, and even more precisely
for the excited states. The accuracy for H$_2^+$ is constrained to
$\ordsim$~1~mHartree for the ground state and less than that for excited states
due to truncation in the angular basis set. When employing a linear
parametrization, a considerably larger number of Hermite-basis functions would be required to reach the same accuracy.

\begin{figure}
	\centering
	\includegraphics[width=\linewidth]{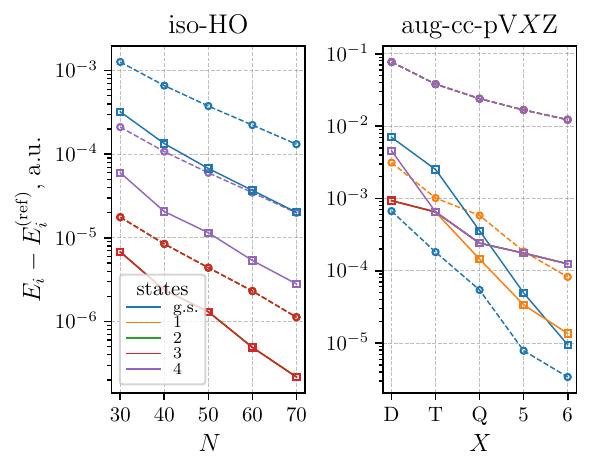}
	\caption{Convergence of the ground (g.s.) and four lowest excited electronic states of hydrogen atom plotted against the size of the radial basis, for the three-dimensional isotropic harmonic oscillator (iso-HO) and atomic-orbital (aug-cc-pV$X$Z) basis sets. Results obtained with fixed-linear (circles, dashes) and normalizing-flow (solid, squares) parameterizations are compared. Since the four excited states of the hydrogen atom are degenerate, this results in indistinguishable and overlapping error values in some of the plots.}
	\label{fig:h}
\end{figure}

Normalizing flows also improved the performance of more common basis sets used in
electronic structure computations, such as the basis set of isotropic
three-dimensional harmonic oscillator (iso-HO) and the augmented
correlation-consistent aug-cc-pV$X$Z ($X$=D, T, Q, 5, 6) atomic orbital basis
sets from Dunning's family~\cite{Dunning:JCP90:1007, Kendall:JCP96:6796}.
Results for hydrogen atom are presented in \autoref{fig:h}, illustrating that the
iso-HO basis set, while inherently well-suited and fast-converging for the
problem, still exhibits a notable enhancement in accuracy when composed with a
normalizing flow.
Although the original atomic-orbital basis set demonstrates quicker ground-state
convergence compared to that composed with a normalizing flow, the latter yields
results that are more accurate and fast-converging for excited states.
The lack of improvement in the ground-state energy by normalizing flow is likely due to the nature of atomic orbital basis sets, which consist of Gaussian functions with exponents carefully optimized for the ground state of an atom.
This specific optimization likely accounts for ineffectiveness of normalizing flow in improving the ground electronic energy.

\begin{figure}[!ht]
	\centering
	\includegraphics[width=\linewidth]{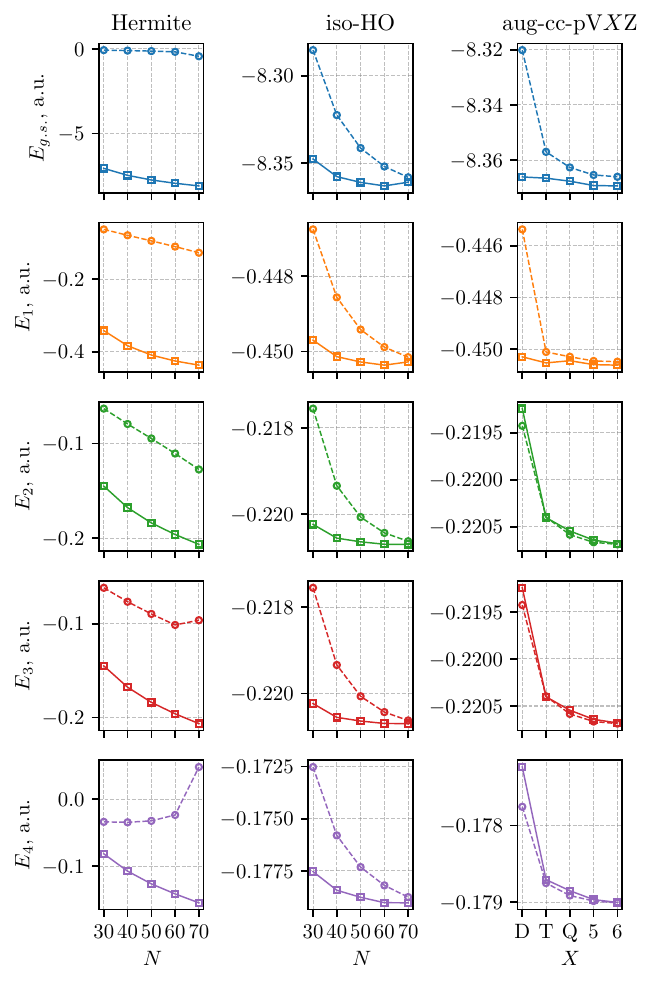}
	\caption{Convergence of the ground (g.s.) and four lowest excited electronic energies of carbon atom plotted against the size of the radial basis $N$ (or $X$), for the Hermite, three-dimensional isotropic harmonic oscillator (iso-HO) and atomic-orbital (aug-cc-pV$X$Z) basis sets. Results obtained with optimized-linear (circles, dashed lines) and normalizing-flow (squares, solid lines) parameterizations are compared.
		The energies are calculated using the single-active-electron approximation with the Coulomb potential obtained from a quantum chemical calculation and neglecting the exchange-interaction potential.}
	\label{fig:catom}
\end{figure}

In \autoref{fig:catom}, we show the convergence for the ground and four lowest
excited electronic states of carbon atom with the number of radial basis
functions, as calculated using the normalizing-flow approach and an optimized-linear parametrization.
The convergence is plotted for different basis sets,
including the Hermite basis set, the isotropic three-dimensional harmonic
oscillator basis set, and augmented correlation-consistent atomic orbital basis
sets. 
The data clearly shows that the normalizing-flow approach generally
improves the accuracy of results for all states across different basis sets,
compared to optimized-linear mapping. However, an exception can be noted in the
case of atomic-orbital basis sets. While normalizing-flow parametrization
leads to faster convergence for the ground and first excited states, it exhibits
comparative or slightly inferior accuracy for the remaining excited states,
particularly with smaller basis sets, as compared to an optimized-linear mapping. We
also noticed some convergence issues of the training procedure with a larger
number of basis
functions, which can be attributed to the limitations in the accuracy of numerical
integration with Hermite and Laguerre quadratures used for calculations across all basis sets.

\end{document}